\documentclass[review]{elsarticle}

\usepackage{lineno}
\usepackage{algorithm}
\usepackage{algpseudocode}
\usepackage{moreverb}
\usepackage{graphicx}
\usepackage{amsmath}
\usepackage{soul}
\usepackage[caption=false]{subfig}
\usepackage{multirow}
\usepackage[table,xcdraw]{xcolor}

\modulolinenumbers[5]

\journal{Journal of \LaTeX\ Templates}

\begin{document}

\begin{frontmatter}

\title{A Survey on Contemporary Computer-Aided Tumor, Polyp, and Ulcer Detection Methods in Wireless Capsule Endoscopy Imaging}

\author{Tariq Rahim, Muhammad Arslan Usman, Soo Young Shin*}
\address{Kumoh National Institute of Technology, Gumi, Gyeongbuk 39177, Republic of Korea}

\author{tariqrahim@kumoh.ac.kr, arslanusman@msn.com, wdragon@kumoh.ac.kr}

\cortext[mycorrespondingauthor]{Soo Young Shin}
\ead{wdragon@kumoh.ac.kr}

\begin{abstract}
Wireless capsule endoscopy (WCE) is a process in which a patient swallows a camera-embedded pill-shaped device that passes through the gastrointestinal (GI) tract, captures and transmits images to an external receiver. WCE devices are considered as a replacement of conventional endoscopy methods which are usually painful and distressful for the patients. WCE devices produce over 60,000 images typically during their course of operation inside the GI tract. These images need to be examined by expert physicians who attempt to identify frames that contain inflammation / disease. It can be hectic for a physician to go through such a large number of frames, hence computer-aided detection methods are considered an efficient alternative.   
\par
Various anomalies can take place in the GI tract of a human being but the most important and common ones and the aim of this survey are ulcers, polyps, and tumors. In this paper, we have presented a survey of contemporary computer-aided detection methods that take WCE images as input and classify those images in a diseased/abnormal or disease-free/ normal image. We have considered methods that detect tumors, polyps and ulcers, as these three diseases lie in the same category. Furthermore bleeding inside the GI tract may be the symptoms of these diseases; so an attempt is also made to enlighten the research work done for bleeding detection inside WCE. Several studies have been included with in-depth detail of their methodologies, findings, and conclusions. Also, we have attempted to classify these methods based on their technical aspects. This paper also includes a potential proposal for joint classification of aforementioned three diseases. 
\end{abstract}


\end{frontmatter}

\section{Introduction}
Physical examination for diagnosing specific diseases in the small intestine leads physicians to suffer from a hectic and time-consuming process. This has led to the development of wireless capsule endoscopy (WCE) technology offering advantages to gastroenterologists for the diagnosis and examination of patients without the execution of complicated surgical operations. Although WCE technology can be used for inspection and assessment of the entire digestive system as an advantage, it can still become a time-consuming procedure, as the complete endoscopic process takes 8 hours, generating 60,000 frames on average. So, an early detection of diseases such as tumors, polyps, and ulcers, is important for widening the scope of treatment options.  
\par
In 1989, a research group in Baltimore invented WCE that is later commercialized it through Given-Imaging Inc.\cite{Givenimaging.com,lin2005blinded}. Table 1 summarizes the companies producing WCE systems, certified by the Food and Drug Administration (FDA); these include MiroCam, IntroMedic Company Ltd., Seoul, Korea; Endocapsule, Olympus America, Inc., Center Valley, Pennsylvania; and PillCam SB2, Given Imaging, Ltd., Yoqneam, Israel. The addition of new technology such as OMOM® Jinshan science of technology capsule offers a good field of view along with good resolution of images still unapproved by FDA \cite{eliakim2017see}. The importance of WCE device is noticeable as WCE allows the small bowel exploration, where significant fatal diseases can be explored outside and inside the small bowel that makes the necessity of an early automatic detection system. Conventional endoscopic technique i.e. gastroscopy and colonoscopy are opted for diagnosis of GI tract still \cite{usman2016detection}, major portion that includes the colon and small intestine remain inaccessible. 
\begin{table}[H]
	\renewcommand{\arraystretch}{1.2}
	\caption{FDA/non-FDA-approved wireless capsule systems and specifications.}
	\resizebox{11.80cm}{!}{
	\begin{tabular}{|c|c|c|c|c|c|c|c|c|}
		\hline
		\textbf{\begin{tabular}[c]{@{}c@{}}WCE company/\\  Capsule\end{tabular}}                                           & \textbf{\begin{tabular}[c]{@{}c@{}}Size\\ “mm"\end{tabular}}                   & \textbf{\begin{tabular}[c]{@{}c@{}}Weight\\ “g"\end{tabular}} & \textbf{\begin{tabular}[c]{@{}c@{}}Field of view\\ Angle\end{tabular}} & \textbf{\begin{tabular}[c]{@{}c@{}}Frames \\ per second\end{tabular}} & \textbf{\begin{tabular}[c]{@{}c@{}}Battery life\\ “hours"\end{tabular}} & \textbf{\begin{tabular}[c]{@{}c@{}}Resolution\\ “pixels"\end{tabular}} & \textbf{Communication}                                                           & \textbf{\begin{tabular}[c]{@{}c@{}}Price/ \\ capsule\end{tabular}} \\ \hline
		\textbf{\begin{tabular}[c]{@{}c@{}}EndoCapsule; Olympus America,\\ Inc., Center Valley, Pennsylvania\end{tabular}} & \textbf{\begin{tabular}[c]{@{}c@{}}Length: 26\\ Diameter: 11\end{tabular}}     & \textbf{3.50}                                                 & \textbf{$145^\circ$}                                                   & \textbf{2 fps}                                                        & \textbf{8 hours or longer}                                              & \textbf{512 $\times$ 512}                                              & \textbf{\begin{tabular}[c]{@{}c@{}}Radio frequency\\ communication\end{tabular}} & \textbf{\$500}                                                     \\ \hline
		\textbf{\begin{tabular}[c]{@{}c@{}}PillCam SB3; \\ Given Imaging,\\ Ltd., Yoqneam, Israel\end{tabular}}            & \textbf{\begin{tabular}[c]{@{}c@{}}Length: 26.2\\ Diameter: 11.4\end{tabular}} & \textbf{3.00}                                                 & \textbf{$156^\circ$}                                                   & \textbf{2 fps or 2-6 fps}                                             & \textbf{8 hours or longer}                                              & \textbf{340 $\times$ 340}                                              & \textbf{\begin{tabular}[c]{@{}c@{}}Radio frequency\\ communication\end{tabular}} & \textbf{\$500}                                                     \\ \hline
		\textbf{\begin{tabular}[c]{@{}c@{}}PillCam SB2EX;\\  Given Imaging\end{tabular}}                                   & \textbf{\begin{tabular}[c]{@{}c@{}}Length: 26\\ Diameter: 11\end{tabular}}     & \textbf{2.8}                                                  & \textbf{$156^\circ$}                                                   & \textbf{2 fps}                                                        & \textbf{12 hours or longer}                                             & \textbf{256 $\times$ 256}                                              & \textbf{\begin{tabular}[c]{@{}c@{}}Radio frequency\\ communication\end{tabular}} & \textbf{\$500}                                                     \\ \hline
		\textbf{\begin{tabular}[c]{@{}c@{}}MiroCam; Intromedic Co., Ltd.,\\ Seoul, Korea\end{tabular}}                     & \textbf{\begin{tabular}[c]{@{}c@{}}Length: 24\\ Diameter: 11\end{tabular}}     & \textbf{3.3-4.70}                                             & \textbf{$170^\circ$}                                                   & \textbf{3 fps}                                                        & \textbf{11 hours or longer}                                             & \textbf{320 $\times$ 320}                                              & \textbf{\begin{tabular}[c]{@{}c@{}}Human body\\ communication\end{tabular}}      & \textbf{\$500}                                                     \\ \hline
		\textbf{\begin{tabular}[c]{@{}c@{}}PillCam ESO2; \\ Given Imaging\end{tabular}}                                    & \textbf{\begin{tabular}[c]{@{}c@{}}Length: 26\\ Diameter: 11\end{tabular}}     & \textbf{$ <4 $}                                               & \textbf{$169^\circ$}                                                   & \textbf{18 fps}                                                       & \textbf{8 hours or longer}                                              & \textbf{256 $\times$ 256}                                              & \textbf{\begin{tabular}[c]{@{}c@{}}Radio frequency\\ communication\end{tabular}} & \textbf{\$500}                                                     \\ \hline
		\textbf{\begin{tabular}[c]{@{}c@{}}OMOM® Jinshan\\ Science and Technology\end{tabular}}                            & \textbf{\begin{tabular}[c]{@{}c@{}}Length: 27.9\\ Diameter: 13\end{tabular}}   & \textbf{6.00}                                                 & \textbf{$169^\circ$}                                                   & \textbf{2 fps}                                                        & \textbf{6-8 hours or longer}                                            & \textbf{640 $\times$ 480}                                              & \textbf{\begin{tabular}[c]{@{}c@{}}Radio frequency\\ communication\end{tabular}} & \textbf{\$250}                                                     \\ \hline
	\end{tabular}
}
\end{table}
\par
Diseases of the gastrointestinal tract (GI) are caused mainly by persistent or recurring bleed-inducing infections. The extracted frames from a WCE video \cite{usman2017quality}, where a small portion of the small bowel is presented showing a yellowish coloration \cite{usman2017quality} along with the faintly pink color of the small intestine. Whenever bleeding occurs, the tissues become red and the affected regions become light pink and purple.
\par
A typical WCE system comprises of three components: (a) capsule endoscope, (b) sensing pads or a belt for attaching the battery pack and data recorder to the patient, (c) a PC workstation having proprietary software. The WCE device, shown in figures 1 and 2, is a pill-sized swallowable, capsule-shaped electronic device comprised of various internally fixed optical and electronic components. The WCE device comprises a camera covering a definite angle of view reliant on the lens. Figure 1 shows the MiroCam, IntroMedic Company Ltd., Seoul, South Korea, device having an 170 degree of view \cite{usman2016detection}.

\par 
Computer-aided technologies such as computer vision particularly offer services for automatic analysis of WCE videos resulting in the decrease of time taken by physicians for processing and examining videos. Current research has focused on two main issues related to the analysis of endoscopic images, namely, the detection and discrimination of malignant tissue. The former involves the detection of malignant intestinal abnormalities such as tumors, polyps, ulcers etc. mainly caused by bleeding.

\begin{figure}[H]
	\centering
	\includegraphics[width=5.0in,height=5.0in,keepaspectratio]{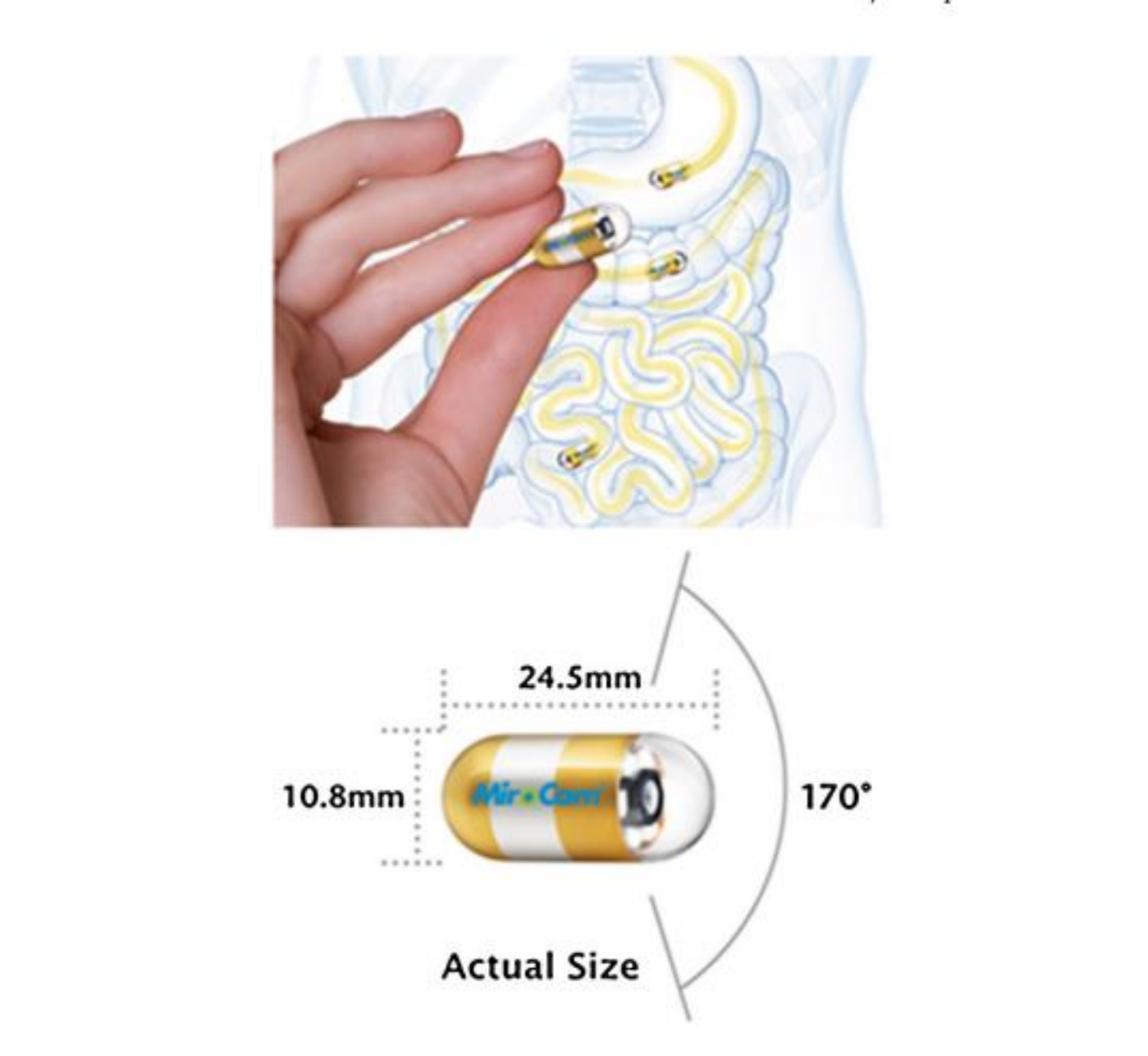}
	\caption{Wireless capsule endoscopy (WCE) device (Capsule MIROCAM by Intromedic Ltd.).}
	\label{Fig1}
\end{figure}
\par 
The motivation behind this survey is to provide a comprehensive analysis of the techniques for detection of tumors, polyps, and ulcers, while considering purely wireless capsule endoscopy (WCE). The focal point is to provide a comparative analysis of the existing techniques, thereby creating possibilities for future research in this specific domain.

\begin{figure}[H]
	\centering
	\includegraphics[width=5.0in,height=4.0in,keepaspectratio]{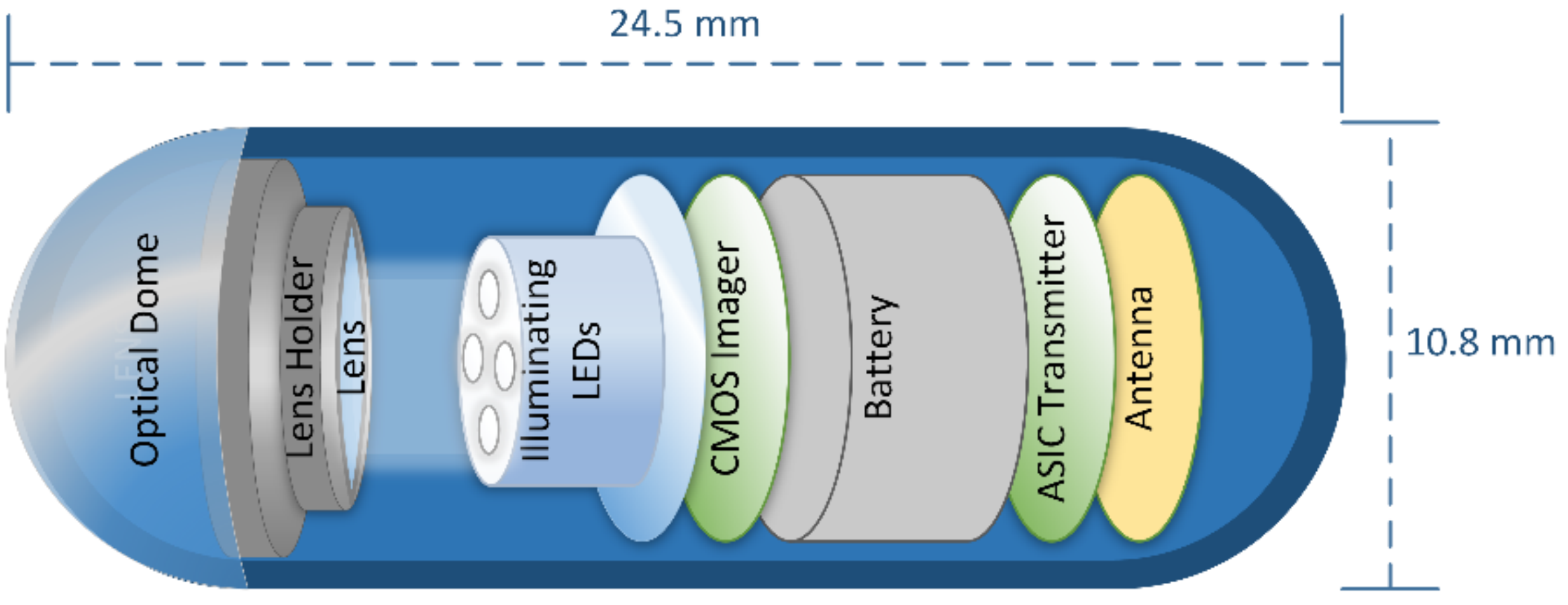}
	\caption{Components of WCE device [3]; CMOS: complementary metal oxide semiconductor, ASIC: application specific integrated circuit. (Dimensions specified for MIROCAM capsule by Intromedic Ltd).}
	\label{Fig2}
\end{figure} 

\par 
The rest of the paper is organized as follows. Section 2 describes the abnormality detection in WCE images. Section 3, 4 and 5 details a comprehensive description of detection for tumor, polyps, and, ulcers in WCE images respectively. Section 6 details the bleeding detection inside WCE images. Section 7 incorporates a Possible future work: Proposed Solution where a cascade approach of neural networks is suggested for detecting tumor, polyp, and ulcer jointly. Section 8 covers a formal discussion in which attempt is made to enlighten different approaches made for detecting tumor, polyp, and ulcer with comparison to the proposed solution with proper references. Finally, the paper is wrapped in section 9, provides a conclusion of the presented survey. Detailed information on the performance of each approach for detection of the respective abnormalities is presented in tabular form. A diagrammatic view is provided that illustrates the technique made for each disease detection in the survey.
\section{Abnormality detection in WCE images}
This section attempts to details the work done for the detection of abnormalities in wireless capsule endoscopic images 
\par 
\par
A unique automatic approach for the detection and localization of the anomalies within the gastrointestinal (GI) from frame sequences of video endoscopic is proposed \cite{iakovidis2018detecting}. The proposed technique classifies the endoscopic frames into abnormal and normal frames by implementing a weakly supervised convolutional neural network (WCNN) architecture followed by Deep Saliency Detection (DSD). The training of the model is done on weakly annotated images i.e. on image-level rather than pixel-level in the RGB color space. For the salient point detection, the deeper hidden layers of WCNN is used to extract the features maps by DSD algorithm. An iterative cluster unification (ICU) algorithm which is based on descriptor i.e. pointwise cross-feature- map (PCFM) extracted by DSD is used for the localization of GI anomalies. This involves both training and testing stage, wherein training phase the ICU is trained with both normal and abnormal training images by clustering their salient points. The images classified by WCNN and detected salient points are received by ICU in the testing phase in a unified cluster form. Publicly available two data sets i.e. MICCAI comprising of 698 images out of which 465 images are used for training and 233 images are used for testing purpose. The second data set used is from KID dataset comprising of 2,352 images. For the classification performance of the proposed WCNN model is evaluated by utilizing different three diverse learning algorithms generating a high performance for both data sets. For the data set obtained MICCAI, the proposed classification method achieved an accuracy, sensitivity, and specificity of 90.90\%, 93.00\%, and 88.50\% respectively, while for the data set obtained from KID, the corresponding accuracy, sensitivity, and specificity of 89.92\%, 92.40\%, and 85.58\% is achieved respectively \cite{iakovidis2018detecting}. For the localization of the anomalies, the area under the curve (AUC) with the best performance value for MICCAI and KID is 84.48\% and 87.70\% respectively is achieved when cluster =2.
\par 
For the abnormalities detection inside WCE images, a technique such as local fuzzy patterns (LFP) is implemented that discriminate frames have normal and abnormal regions \cite{maghsoudi2013detection}. The proposed method is based on two methods i.e. color, geometry (invariant moments), and texture features (GLCM, Gabor, LBP, and Laws' features). The LBP is used for the extraction of color texture features in four different color channels i.e. red, green, hue, and grayscale for the detection of abnormal and normal regions. To overcome with limitations of LBP, the LFP is implemented to extract features as unlike to LBP, the LFP is handling the idea of partial truth i.e. completely false and completely true. The data set comprised of 44 normal and 100 abnormal for the analysis by maintaining the parameters for both LBP and LFP constant. The proposed approach for abnormalities regions detection inside WCE images generated results in terms of accuracy, sensitivity, and specificity of 90.00\%, 92.00\%, and 91.00\%.  respectively \cite{maghsoudi2013detection}.
\par 
Furthermore, an automatic technique where splitting of WCE image into several patches is done for the extraction of features. The speeded-up robust features (SURF) are extracted from each pixel of the image instead of utilizing the blob nature of SURF algorithm in CIELAB and “YCbCr" color space. To overcome the manual crafting of the feature vector of SURF interest points and lessening the burden of a large number of WCE images crafting, CNN is implemented that learn patterns of textural color and extract features from different regions of WCE image \cite{sekuboyina2017convolutional}. The CNN gets the patches instead of the whole image and to avoid redundant patched information leading to overfitting, the WCE frame is spliced into nonoverlapping patches based on a percentage of malignant or benign pixels. Using a trained model, classification is achieved followed by assignment of each pixel to their respective patch. The proposed approach for the training data achieved sensitivity and specificity of 92.00\% and 97.00\% respectively while for testing data the sensitivity and specificity achieved is 21.00\% and 98.00\% respectively in CIELAB color space. The overall area under the curve (AUC), sensitivity, and specificity for all considered diseases in the specific study are 79.61\% to 83.00\%, 71.00\% to 90.00\%, and 72.00\% to 75.00\% respectively outperforming the compared approaches \cite{sekuboyina2017convolutional}.
\par 
Moreover, for the discrimination of pathogenic endoscopic frames associated to ulcer within the gastrointestinal tract (GT), a unique method based on the Bidimensional Ensemble Empirical Mode Decomposition (BEEMD) \cite{charisis2010abnormal} is implemented for the extraction of their Intrinsic Mode Functions (IMFs). The BEEMD \cite{wu2009multi} which is an extension of EEMD that pursue and reveal any kind of inherited intrinsic oscillatory modes, named IMF by remaining in the time domain. The primary step for the proposed approach is the specification of the region of interest (ROI) which is manually done by cropping in the proposed study followed by a selection of RGB color space and channels extraction.  Due to the nature of WCE images generation, noises are eliminated by applying BEEMB on each channel of color followed by decomposition of each image into eight IMFs per channel. Then for the quantification and extraction of textural features, lacunarity analysis (Lac) is applied on each channel of the RGB color space. The lacunarity window size curve is generated to examine the capability of classification of WCE images. For the classification purpose, discriminant analysis-based classification, and support vector machine (SVM) are used. The total data set comprised of 80 images, out of which 40 pathogenic and 40 normal images. For both classifiers, the proposed approach \cite{charisis2010abnormal} is achieving high performance in the green channel, while for overall performance as benchmarked, the SVM outperforms by achieving an accuracy, sensitivity, and specificity of 95.75\%, 95.00\%, and 96.50\% respectively.
\par 
Furthermore, for the detection and classification of gastrointestinal tract diseases in wireless capsule endoscopy (WCE) images, a novel technique of combining of deep convolutional neural network (DCNN) and geometric features is proposed. A novel approach i.e. contrast-enhanced colorcolour features (CECF) and feature extraction via CNN is proposed for the extraction of lesion region \cite{sharif2019deep}. The CECF gets along with two sub steps: 1) enhancing the local contrast of infectious region via hybrid contrast stretching method that comprised of median filter and top-bottom hat filtering; extraction of color features via Hue, Saturation, Variation (HSV) transformation that will be used for similarity measures among pixels of HSV color image. The stretching phase consisting of top-hat filter enhance foreground contrast while bottom hat filter enhances background contrast removing unnecessary information and controlling global contrasts of the image. For the lesion detection CIELAB color transformation is done on the median filtered images for the extraction of color features such as variance, standard deviation, mean, kurtosis, and skewness. These color features are then integrated in a matrix and mean deviation is computed by which weighted value is generated for making a cluster of the features extracted. Lastly, a conditional probability dependent threshold value is generated that is used for segmentation. Extraction of geometric and CNN features from improved RGB and segmented images is done. Pretrained model of VGG16 and VGG19 are used for the extraction of deep features from WCE images followed by $3\times3$ max-pooling operation for the selection of features with higher values and information. Then a unique method i.e. Euclidean Fisher Vector (EFV) is proposed for the fusion of features extracted from both trained models. In the EFV method, fisher vector-based similar features are ordered into one separate cluster and other features are ordered into the different cluster. EFV method provides a unique feature vector which is fused parallel latter in one matrix. Following, the conditional entropy strategy is implemented on a fused vector for the selection of best features where features greater than 0.4 are selected for final classification. For the classification, the final selected features become the input of Fine K-Nearest Neighbor (FKNN) \cite{larose2014discovering}. The proposed method is implemented on total of 4500 frames, out of which 2000 bleeding and health frames respectively, while 500 are ulcer images. For the proposed fused features approach, the KNN generated a highest performance in terms of accuracy, sensitivity, precision, and specificity of 99.42\%, 100\%, 99.51\%, and 100\% respectively \cite{sharif2019deep}. 
\par 
Furthermore, for the automatic detection of patterns in wireless capsule endoscopy (WCE) images, an integrated image processing with recognition analysis is presented. In particular, the method performed consists of a selection of techniques such as image registration, segmentation, local-global (L-G) graphs, region synthesis, and neural nets merged into a synergistic fashion for obtaining the desired outcomes \cite{bourbakis2005detecting}. For image registration phase, genetic algorithm optimization tactic is applied for finding the optimal registration parameter within the global search space. The region-based segmentation is achieved by applying Fuzzy-like Reasoning Segmentation (FRS) technique which has three phases i.e. smoothing, edge detection, and segmentation. The smoothing phase will eliminate noise via low pass filter in a selective manner that preserve edges of images location. The segmentation phase comprises of flood fill operations for the interior points of images that decide based on color similarity in the adjacent pixels to be filled or not. For the accurate representation and interest points, the L-G graph technique is adopted avoiding non-linear graph matching function. By integration the FRS with L-G graph the recognition accuracy of object can be improved. A neural nets-based recognition developed by the ATL research lab is used for the detection purpose and the results are showed in a figure wise in the proposed work lacking the performance metrics \cite{bourbakis2005detecting}. 
\par 
For addressing the problem related to gastrointestinal walls that deforms in erratic way, a tracking technique for wireless capsule endoscopy (WCE) images using triangular constraints via an affine transformation is proposed. The motivation is that the location of the target can be determined in the successive frames transition by using affine transformation \cite{yanagawa2017abnormality}. The proposed approach after tracking the abnormal region both in backwards and forwards direction involves three stages. First, matching within successive frames for feature points in the abnormal (target) and surrounding (supporters) are done via SIFT / SURF and Kanade-Lucas-Tomasi (KLT) method \cite{tomasi1991tracking} respectively. Second is the estimation of the target position is done based on voting process via affine matrix which is computed from triplet supporters. Adjustment of the relative location of the target is done by mapping of triangles among current and successive frames. Third, precise position determination of abnormal region is done, and to avoid the error induced by the movement of successive frame; color information of the target is utilized. The proposed abnormality tracking approach is applied on 120 sequences at frame rate 2 fps having eight major abnormalities. The proposed method \cite{yanagawa2017abnormality} for abnormality tracking outperformed other efficient tracking algorithm such as mean-shifted technique \cite{collins2003mean} and the multi-domain network (MDNet) \cite{nam2016learning} technique. 
\par 
Table 2 summarizes the survey for abnormality detection in WCE images in terms of the performance of the respective methods. The accuracy, specificity, sensitivity, and precision for the abnormality detection by the respective approaches are indicated in the table.
\begin{table}[H]
	\renewcommand{\arraystretch}{0.65}
	\caption{Efficacy of existing methods for Abnormality detection in WCE images.}
	\resizebox{11.30cm}{!}{
	\begin{tabular}{|c|c|c|c|c|}
		\hline
		\textbf{Ref No.}  & \textbf{\begin{tabular}[c]{@{}c@{}}Accuracy\\ (\%)\end{tabular}} & \textbf{\begin{tabular}[c]{@{}c@{}}Specificity\\ (\%)\end{tabular}} & \textbf{\begin{tabular}[c]{@{}c@{}}Sensitivity\\ (\%)\end{tabular}} & \textbf{\begin{tabular}[c]{@{}c@{}}Precision\\ (\%)\end{tabular}} \\ \hline
		\textbf{{[}6{]}}  & 90.90                                                            & 88.50                                                               & 93.00                                                               & \textbf{NO}                                                       \\ \hline
		\textbf{{[}7{]}}  & 90.00                                                            & 91.00                                                               & 92.00                                                               & 91.70                                                             \\ \hline
		\textbf{{[}8{]}}  & \textbf{NO}                                                      & 75.00                                                               & 90.00                                                               & \textbf{NO}                                                       \\ \hline
		\textbf{{[}9{]}}  & 95.75                                                            & 96.50                                                               & 95.00                                                               & \textbf{NO}                                                       \\ \hline
		\textbf{{[}11{]}} & 99.42                                                            & 100                                                                 & 100                                                                 & 99.51                                                             \\ \hline
		\textbf{{[}13{]}} & \textbf{NO}                                                      & \textbf{NO}                                                         & \textbf{NO}                                                         & \textbf{NO}                                                       \\ \hline
		\textbf{{[}14{]}} & \textbf{NO}                                                      & \textbf{NO}                                                         & \textbf{NO}                                                         & \textbf{NO}                                                       \\ \hline
	\end{tabular}
}
\end{table}

\section{Tumor detection in WCE images}
This section tries to describe the work done for the detection of tumors in wireless capsule endoscopic images by pointing out both machine learning and other image processing approaches. 
\par 
In order to characterize WCE images into abnormal, normal and tumor categories, an adaptive neuro-fuzzy interface system (ANFIS) is implemented \cite{alizadeh2017detection}. The technique subsequently segregates images based on statistical analysis by following two steps i.e. extraction of the features and classification of the images based on 32 features. The features also include four statistical factors, i.e. contrast, homogeneity, correlation, and energy, obtained from co-occurrence metrics. Minimum redundancy among the features is achieved via mutual information selection of the features with maximum dependence on the target class generating a classification accuracy, specificity, and sensitivity of 94.20\%, 96.27\%, and 94.16\% respectively.
\par 
Furthermore, a two-step automatic tumor detection procedure i.e. Region of Interest selection and classification is proposed inside WCE images. The detailed pipeline of the proposed approach comprises pre-processing, segmentation, feature extraction and classification \cite{vieira2019automatic}. Color conversion from “RGB" to “CIELab" color space constitutes a pre-processing stage in which a and b channels selected for segmentation purpose. In the first step, ROI is detected automatically via a segmentation module to find area having abnormal tissue partially. An algorithm such as Maximum a Posteriori (MAP) approach and the Expectation-Maximization (EM) for measuring the parameters of a Multivariate Gaussian Mixture Model (GMM) is used for segmentation. For convergence acceleration of the Expectation-Maximization (EM), a modified version of the Anderson algorithm is proposed. For the feature extraction histogram-based, statistical features such as variance, kurtosis, mean, and entropy are extracted. The second step utilizes the ensemble learning (EL) system grounded on SVMs where Bagging tactic is utilized for obtaining a global classification. In the case of ES, the variety which ensures that per member of the ensemble is tuned for a subset of the training set can be obtained by using bootstrapping where the training data is randomly sampled with replacement. Two GMM are utilized i.e. one for normal data clustering (M clusters) and one for clustering abnormal data (N clusters) and combined into $M\times N$. Every subset is used to train a Least Squares-SVM classifier. At this point, the gating network i.e. MLP is trained for decoupling the collective maximization of both modules and same process done for testing step. The data set consisted of total 3936 frames out of which 3000 normal and 936 are pathological frames and are trained on WEKA for each ensemble elements (SVMs). The proposed technique of extracting features via ensemble classifier \cite{vieira2019automatic} generated a high accuracy, specificity, sensitivity, and area under the curve (AUC) of 96.30\%, 96.70\%, 95.00\%, and 95.10\% respectively that also outperforms features extracted by Wavelets and Curvelets transforms in more than 5\%.  
\par 
To discriminate tumor from normal images, a unique set of features based on the texture that assimilates a multi-scale curvelet is operated in combination with fractal technology \cite{liu2016detection}. Executing a transformation known as inverse curvelet of certain scales,  texture descriptors are computed of second and high order moments for different color channels.  Genetic algorithm (GA) and support vector machine (SVM) is implemented in order to optimize set of features and classification respectively. The results after benchmarking with other approaches showed better performance in terms of sensitivity, specificity, and accuracy of 97.80\% and 96.70\%, and 97.30\% respectively. Moreover, out of those 89 clinical patients under the study, 16.8\% (15) were diagnosed \cite{liu2016detection}.
\par 
Due to the diverse appearances and shapes of images captured by capsule endoscopy, a unique textural feature based on wavelet and the multi-scale local binary pattern is implemented in \cite{li2011computer}. Three classifiers i.e. multi layer perception neural network (MLP), k-nearest neighbor, and SVM are implemented to facilitate classification and distinction of the diverse tumor images. The data set consists of 1200 images obtained from ten patients are used for this study resulting in detection accuracy, sensitivity, and specificity of 90.50\%, 92.33\%, and 88.67\%, respectively.   
\par 
A wavelet decomposition called color wavelet co-variance (CWC) is implemented in \cite{karkanis2003computer}, that extract color features for the detection of tumor in endoscopic videos. Based on second-order textural computation of the co-variances, features are determined followed by optimal subset selection of features via the selected algorithm. Classification is achieved by linear discriminant analysis (LDA) in support of the proposed algorithm on color colonoscopic videos generating a high specificity and sensitivity of 97.00\% and 90.00\%, respectively. 
\par 
Furthermore, an integrated approach of using discrete wavelet transform (DWT) and singular value decomposition (SVD) algorithm are implemented for tumor detection \cite{faghih2016singular}. Using DWT diversities in the appearance and illumination environment (which generally change with respect to time) are coped with. Finally, SVM is applied for classifying the WCE images to a data set constituting 400 normal and 400 abnormal WCE images. The integrated approach achieved a sensitivity, specificity, and accuracy of 94.00\%, 93.00\%, and 93.50\%, respectively. 
\par 
Furthermore, a computer-aided diagnostic technique for detection of small bowl tumor inside WCE images is presented in \cite{li2009small}. The work take advantages of integrating wavelet transform and textural feature called local binary pattern as extractor in order to discriminate normal regions from tumor regions. Implementing uniform local binary pattern  for multi-resolution and rotation invariant textural features are extracted followed by three level of DWT on each channel of WCE image in “RGB" and “HSI" color spaces. For classification, support vector machine (SVM) is implemented on data set of 300 images obtained by PillCam SB2 yielding an accuracy, specificity and sensitivity of 96.67\%, 96.00\% and 97.33\% in HSI color space \cite{li2009small} showing that efficiency of usage of wavelet based local binary pattern approach.  
\par 
An integrated computer-aided scheme combining the textural features, uniform local binary pattern, wavelet, and SVM is implemented for the detection of tumors in the digestive tract \cite{li2012tumor}. The local binary pattern is a descriptor of a local texture that defines the intensity distribution and is unaffected by illumination variances and offers multi-resolution of WCE images. SVM  used for classification; nevertheless, the selected features do not guarantee high performance in terms of classification. Therefore, two additional feature selection tactics were implemented namely, sequential forward floating selection based on SVM (SVM-SFFS) and recursive feature elimination based on SVM (SVM-RFE) \cite{jain1997feature}. In the SFFS approach, some steps are performed in reverse subsequent to the forward steps \cite{guyon2002gene}, while in SVM-RFE, backward feature removal is used to rank the features based on weight, which is a function of the support vectors \cite{li2012tumor}. This two-feature selection helped in improving the detection capability, as the features were more refined. The suggested technique was validated by conducting a wide range of experiments, resulting in an accuracy, sensitivity and specificity of 92.40\%, 88.60\% and 96.20\% for recognition of the tumors in WCE images.
\par 
As mentioned, visualization of the entire GI tract is important and can be achieved by WCE. Human textural perception is dependent on the multi-scale investigation of patterns and was simulated using multi-resolution tactics \cite{barbosa2009automatic}. Moreover, the covariance of textural descriptors was successfully used in the classification of colonoscopy videos. A frame classification pattern that is dependent on the statistical textural features obtained by applying the discrete curvelet transform (DCT) domain was suggested [16]. The DCT domain is a newly introduced multi-resolution tool. To address the limited directional sensitivity of DWT, DCT was used as it offers high directional sensitivity in many directions and is anisotropic in nature. These properties make DCT the best option for dealing with pattern complexities such as texture. The classification features obtained from the textural co-variance at multiple angles were applied to the proposed color curvelet co-variance scheme. The mean and standard deviation of the coefficients of DCT was the best descriptors of the textural features in this case \cite{barbosa2009automatic}. The classification is achieved by implementing an (MLP) neural network, resulting in a sensitivity of 97.20\% and specificity of 97.40 \%. 
\par 
Due to the variations in the appearance, luminance, texture of the images of the intestine and variation of the shape among normal tissues and hemorrhages computerized detection technique becomes a hectic task.  Owing to the astonishing number of image analysis errors, the convolutional neural networks (CNNs') model is
preferred and has been used for the recognition of intestinal hemorrhage \cite{li2017convolutional}.  CNN models include LeNet, AlexNet, GoogLeNet, and VGGNet that offer variation in the format of the input, depth, and modules are explored in this study. Data augmentation using various transformations has been used to prevent over fitting, which is a severe problem when dealing with CNN models. The correlation between the image quality and detection accuracy of intestinal hemorrhage was also investigated. The CNN model is applied on a dataset comprising of 1300 of hemorrhage images and 40,000 normal images resulting highest precision for AlexNet of 98.06\% and F-measure for VGG-Net of 98.87\% showing the validity of the different models \cite{li2017convolutional}.
\par 
A textural feature extraction technique followed by SVM in order to classify tumor/abnormal from normal images. The pre-processing is done via median filtering to denoise the images followed by DWT as feature extractors due it's multi-orientation characteristics. The textural features (mean and energy) are computed for diagonal components as a mean, while energy is considered for horizontal and vertical parts. These features are used for training the support vector machine recursive future elimination (SVMRFE) classifier that has the advantage of avoiding selection duplicate values in the extracted features \cite{Ashokkumar2014AutomaticDO}. This method was applied to a small number of images, with 100\% accuracy in detecting tumors in WCE images.
\par 
A novel approach that exploits the motion of animals i.e., mice, which is archetypal of neuroscience is used to track and detect frames having tumors, bleeding pixels, and other abnormal images. Quantification of the kinematics of running rodents is necessary in case of high frame rates (250Hz). For this purpose, an automated technique is necessary to track the paws of animals, where the method constitutes the following steps: color-based segmentation, followed by classification using SVM plus neural network (NN) \cite{Maghsoudi2018FeatureBF}. By coupling the kinematic features of running rodents and textural features, further classification and identification are performed to ensure precise analysis of the labeled paws. For detecting the frames with tumors in the dataset, features such as the geometry are extracted from the central region followed by Gabor filter for segmenting the features.  Extraction of 22 gray level co-occurrence matrix (GLCM) features, 4 statistical features (mean, kurtosis, variance, and skewness), and seven invariant moments are done, leading to 990 features.  Moreover, extraction of 75 Law’s features, including the entropy, skewness, variance, mean, and kurtosis of 15 images is achieved by convoluting the images with Law’s masks.  In addition, extraction of 88 features from GLCM at four altered angles plus the calculation of seven invariant moments is done, resulted in a total of 1160 extracted features. Due to the use of dissimilar extraction techniques, normalization is performed. For recognition of the frames with tumors, thirty high-discriminant features are obtained by applying the Fisher test. Thereafter, for the classification of the WCE frames, multi-layer perceptron (MLP) neural network \cite{Mohapatra2012LymphocyteIS} is implemented.
\par
After detection of frames, sub-division of the images is performed and the local binary patterns LBP1 and LBP2 are used to extract 110 textural features, comprising 74 from LBP1 and 36 from LBP2. Then, extraction of the GLCM features at four multi-angles, and mean feature extraction is done making a total of 202 features. For seven samples, 21 masks having two dimensions are obtained by implementing Law’s kernels for five features,  generating 105 features. The kurtosis, skewness, mean, variance, and entropy features are extracted via implementing eight Gabor filters at four angles and two frequency levels to the sub-images, resulting in 50 features in this step. The HSV color space is commonly used for the detection of objects in medical imaging. The kurtosis, skewness, mean, and variance were extracted for five color channels, thus generating a total of 381 features. The aforementioned normalization was performed, resulting in a reduction of features to 30, and subsequent implementation of the Fisher test enabled distinguishing of the chosen features into abnormal and normal regions within a frame \cite{barbosa2009automatic}. For distinction of tumors, other abnormalities, and bleeding from normal tissue, NN was used to classify the regions into normal and abnormal. Increasing the features while applying the Fisher test produced a direct relationship: a sensitivity of up to one and specificity of 0.928. The second technique for recognizing regions with abnormalities gave an accuracy for tumor, bleeding, and abnormality detection of 0.9092, 0.9747, and 0.9461, respectively, and a corresponding precision of 0.8945, 0.9465, and 0.8568; sensitivity of 0.9273, 0.9733, and 0.9671; and specificity of 0.9029, 0.9754, and 0.9381, respectively. For the recognition of tumors only, the second technique gave a specificity and sensitivity up to 0.9011 and 0.9263, respectively, for higher features \cite{Maghsoudi2018FeatureBF}.
\par 
Textural analysis and representation are preferable while detecting tumors in the small intestine using endoscopic images. In the specific approach, wavelet transform is used to select the bands having the most important textural information for different color channels. From the chosen wavelet coefficients for each color channel, the feature set is computed as a co-occurrence matrix \cite{Barbosa2008DetectionOS}. These statistical descriptors are largely based on human insight and judgment of textures. A MLP neural network called back propagation learner having 24 input neurons, and 6, 8, or 12 hidden layers are used. For classification, two output layers are utilized to categorize the images into normal and tumor. The feasibility of the proposed technique is proven by application to capsule endoscopic data obtained from humans, where the sensibility and sensitivity achieved are  98.70\% and 96.60\%, respectively \cite{Barbosa2008DetectionOS}.
\par 
In order to detect tumors and colonic polyps in WCE images, an NN-based approach is presented for extraction of the textural features. Extraction of the features is based on scale invariant feature point (SIFT), which is used not only for a single key point but also for key points in the neighborhood \cite{Sindhu2017AutomaticDO}. The Haralick texture features are extracted for individual $16\times 16$ patches near key points. Combination of SIFT and the Haralick texture features resulted in successful extraction of the features. For classification neural network (NN) is trained on extracted features in order to accurately detect tumors or colonic polyps. The proposed technique achieved an overall accuracy level of 97.50\% with a specificity and sensitivity of 93.40\% and 98.80\%, respectively, for the detection of polyp vs. tumor vs. normal images. In the case of tumor vs. normal detection, the accuracy was 95.10\%, while the specificity and sensitivity were 92.10\% and 96.50\%, respectively. 
\par 
A similar approach is used for detecting several inflammatory bowel diseases (IBDs), including Crohn’s disease, ulcers, tumors, and other abnormalities. Similar to the former approach \cite{Sindhu2017AutomaticDO}, extraction of the features is computed not only for a single key point but also for key points in the neighborhood via integrated SIFT and Haralick texture features implementation. A GLCM matrix having dimensions of $16\times 16$ is generated, and an MLP neural network is then used as a classifier using supervised learning \cite{Sindhu2017ANM}. The results obtained from this method prove the feasibility of the proposed technique, with accuracy values of 96.90\%, 96.40\%, 98.90\%, and 97.40\% for the detection of Crohn’s disease, ulcers, polyps, and tumors, respectively. In addition, the overall accuracy of the proposed technique achieved is 88.60\%.
\par 
Moreover, an automatic small bowel tumor detection is proposed by extracting features in CIELab color space. Discrimination of tumor from normal tissues is done via color histogram information in which light-saturated region is categorized as a tumor \cite{vieira2015automatic}. Maximum a Posteriori approach by using the Expectation-Maximization algorithm is used for the segmentation process, where parameters of a Multivariate Gaussian Mixture Model (GMM) are estimated. To lessen the influence of light in medical images, CIELab color space is used where L channel is discarded remaining a and b channels. Five histogram-dependent statistical features i.e. mean, variance, entropy, kurtosis and the value reached by the cumulative function for 95.00\% of the data are extracted. A total of 3200 frames are used for the analysis in which 700 and 2500 frames are a tumor and normal frames respectively. For classification Multilayer Perceptron neural network (MLP) and a Support Vector Machine (SVM) with cross-validation of 10-fold are implemented. The proposed detection technique using feature grouping of mean and variance over ROI resulted in accuracy, sensitivity, and specificity of 98.80\%, 98.50\%, and 98.80\% respectively \cite{vieira2015automatic}.

\begin{table}[H]
	\renewcommand{\arraystretch}{0.60}
	\caption{Efficacy of existing methods for Tumor detection in WCE images.}
	\resizebox{11.30cm}{!}{
	\begin{tabular}{|c|c|c|c|c|}
		\hline
		\textbf{Ref No.}  & \textbf{\begin{tabular}[c]{@{}c@{}}Accuracy\\ (\%)\end{tabular}} & \textbf{\begin{tabular}[c]{@{}c@{}}Specificity\\ (\%)\end{tabular}} & \textbf{\begin{tabular}[c]{@{}c@{}}Sensitivity\\ (\%)\end{tabular}} & \textbf{\begin{tabular}[c]{@{}c@{}}Precision\\ (\%)\end{tabular}} \\ \hline
		\textbf{{[}18{]}} & 94.20                                                            & 96.27                                                               & 94.16                                                               & \textbf{NO}                                                       \\ \hline
		\textbf{{[}19{]}} & 96.30                                                            & 96.70                                                               & 95.00                                                               & \textbf{NO}                                                       \\ \hline
		\textbf{{[}20{]}} & 97.30                                                            & 96.70                                                               & 97.80                                                               & \textbf{NO}                                                       \\ \hline
		\textbf{{[}21{]}} & 90.50                                                            & 88.67                                                               & 92.33                                                               & \textbf{NO}                                                       \\ \hline
		\textbf{{[}22{]}} & \textbf{NO}                                                      & 97.00                                                               & 90.00                                                               & \textbf{NO}                                                       \\ \hline
		\textbf{{[}23{]}} & 93.50                                                            & 93.00                                                               & 94.00                                                               & \textbf{NO}                                                       \\ \hline
		\textbf{{[}24{]}} & 96.67                                                            & 96.00                                                               & 97.33                                                               & \textbf{NO}                                                       \\ \hline
		\textbf{{[}25{]}} & 92.40                                                            & 96.20                                                               & 88.60                                                               & \textbf{NO}                                                       \\ \hline
		\textbf{{[}28{]}} & \textbf{NO}                                                      & 97.40                                                               & 97.20                                                               & \textbf{NO}                                                       \\ \hline
		\textbf{{[}29{]}} & \textbf{NO}                                                      & \textbf{NO}                                                         & \textbf{NO}                                                         & 98.06                                                             \\ \hline
		\textbf{{[}30{]}} & 100                                                              & \textbf{NO}                                                         & \textbf{NO}                                                         & \textbf{NO}                                                       \\ \hline
		\textbf{{[}31{]}} & 90.92                                                            & 90.11                                                               & 92.63                                                               & \textbf{NO}                                                       \\ \hline
		\textbf{{[}33{]}} & NO                                                               & NO                                                                  & 96.60                                                               & NO                                                                \\ \hline
		\textbf{{[}34{]}} & 95.10                                                            & 92.10                                                               & 96.50                                                               & \textbf{NO}                                                       \\ \hline
		\textbf{{[}35{]}} & 88.60                                                            & 90.70                                                               & 72.50                                                               & \textbf{NO}                                                       \\ \hline
		\textbf{{[}36{]}} & 98.80                                                            & 98.80                                                               & 98.50                                                               & \textbf{NO}                                                       \\ \hline
	\end{tabular}
}
\end{table}
\par 
Table 3 summarizes the survey for detection of tumors in wireless capsule endoscopy images in terms of the performance of the respective methods. The accuracy, specificity, sensitivity, and precision for the detection of tumors by the respective approaches are indicated in the table. 
\section{Polyp detection in WCE images}
The detection of other pathological defects apart from tumors in the digestive tract, such as polyps, ulcers, and Crohn’s disease, is also important for timely diagnosis. A unique technique for the detection of polyps and ulcers in the perforated form in WCE frames is attempted \cite{Karargyris2011DetectionOS}. Polyps and ulcers share the common property of having different perceptible patterns and geometries, making them easily noticeable by experts. However, these abnormalities have different colors and different sharp edges, although converse to ulcers, the color of polyps does not vary from that of the normal digestive tract and wall, which makes it hard to extract color features in this case. Detecting polyps in such a scenario, as well as in WCE images where polyps can occur anywhere, is thus a difficult task. The preprocessing step that includes segmentation should achieve the goals of maintaining the boundary information of the images and extraction of sharp segments, making this preprocessing step important. The log Gabor filter and Susan edge detector are thus used for segmentation and edge detection. The Gabor filter has the property of increased spectral information, Gaussian response and the ability for bandwidth optimization making it the best candidate for segmentation \cite{Field1987RelationsBT}. The smallest uni-value segment assimilating nucleus (SUSAN) edge detector is used to further improve the technique and the preprocessing step. Extraction of the geometric features is performed subsequently to obtain the output of the log Gabor and SUSAN edge process for detection. The color space and its transformation are important features for detecting ulcers in WCE images. HSV is recommended for medical imaging analysis, and transformation of RGB into HSV is preferred. The fuzzy region-growing tactic is the basis of the segmentation structure. Smoothing was used to remove noise from the input image while preserving the locality of edges in the areas where local contrast was inevitably low.  A supervised learning technique i.e., SVM, was used to take advantage of the property of mapping inputs in n-dimensional space by finding the optimum hyperplane for separation of datasets. In addition to extraction of the pattern of ulcers, features such as the statistical texture were also computed \cite{Haralick1973TexturalFF} in word format with 14 values. This word pattern was introduced into the fuzzy support vector machine as an input in order to classify the input pattern \cite{Inoue2001FuzzySV}. Fifty frames of WCE video comprising 40 normal and 10 polyp frames, respectively, were subjected to the algorithm for calculating the sensitivity and specificity, resulting in a high sensitivity of 100\% and low specificity of 67.50\%. The results demonstrate the feasibility of the technique for detecting only polyp and ulcers, independent of tumor detection \cite{Karargyris2011DetectionOS}.
\par 
Furthermore, in another work to detect polyp in WCE images is based on the extraction of features. Frames having the region of none-intreset inside WCE videos are discarded. A sliding window of size $227\times 227$ on the frame is slid from top to bottom and left to right generating window images \cite{Billah2017AnAG}. Using color wavelet (CW) transform, textural features are extracted due to the variation in polyp appearance. The co-occurrence matrix was computed in four different directions in order to find spatial relationships. Additional features like blobs and edges are extracted with a window size of $227\times 227$ using a convolutional neural network (CNN) stimulated by the work done in \cite{Mesejo2016ComputerAidedCO}. Using multi-class SVM that provide sparse, noisy data and a fast linear solver, classification is performed on extracted features from window image.  Compared to other techniques, the proposed technique showed better performance in terms of accuracy, sensitivity, and specificity, with values of 98.65\%, 98.79\%, and 98.52\%, respectively \cite{Billah2017AnAG}.
\par
Furthermore, a computer-aided polyp detection technique is proposed based on the amalgamation of features extracted from color wavelets and convolutional neural network (CNN) \cite{billah2018gastrointestinal}. Due to variance in the shape and size of polyps, multi resolution analysis via wavelet is done for color texture features in the RGB color space. Statistical features such as homogeneity, correlation, entropy, and energy are computed resulting in 144 color wavelet features. The CNN model at the beginning layers of the network extracts primitive features i.e. blobs and edges that are processed further in the deeper layers for obtaining a high level of image features.  All the features extracted from both color wavelet and CNN model are fed to support vector machine (SVM) for the classification purpose. The data set comprised of 14000 images in which 3500 and 10500 images consisted of polyps and non-polyps respectively. The proposed features fusion scheme generated a high performance in terms of accuracy, sensitivity, and specificity of 98.34\%, 98.67\%, and 98.23\% respectively \cite{billah2018gastrointestinal}.
\par 
Moreover, an integrated approach of the Gabor filter and monogenic-local binary pattern (M-LBP) is implemented in order to detect polyp inside 55,000 images obtained from one patient \cite{Yuan2014ANF}. By exploiting the excellent spatial locality and selection of orientation provided by the Gabor filter, multi frequency textural information is extracted. To avoid over-simplification of the local image shapes, M-LBP \cite{Zhang2010MonogenicLBPAN} was used in combination with LBP as a rotation invariant technique for each response image of the Gabor output. M-LBP comprises LBP operator information and a local surface and local phase. Dimension reduction is an important step to deal with a large number of elements and aids in the selection of the features that are the most important for the classification of polyps. Application of a supervised algorithm termed local discriminant analysis (LDA) \cite{Karargyris2011DetectionOS} is performed in order to achieve dimension reduction of the features. For classification, SVM was applied to a set of 872 images, half of which contained polyps and half comprising normal tissue images. The proposed technique resulted in a detection accuracy, sensitivity, and specificity values of 91.43\%, 88.09\%, and 94.78\%, respectively.
\par 
Furthermore, computerized detection of polyps in WCE images is performed based on the unique textural features obtained by combining wavelet and uniform local pattern \cite{li2012automatic}. Using SVM, classification is achieved. The LBP \cite{ojala1996comparative}, which is a textural operator robust to any kind of transformation extended \cite{ojala2002multiresolution} to achieve wider multi-resolution characteristics for textural descriptors. Separation of operator used in \cite{ojala2002multiresolution} is done defined by parameters such as (P) and (R) , that governs quantization of the region in the angular domain and radius of a circle that controls the spatial resolution of this descriptor \cite{li2012automatic}. DWT is implemented for multi-resolution texture analysis at different sub-energy level bands by integrating multiple operators having different values of (P) and (R). During the experiment, cross-validation was performed to avoid the issue of over-fitting \cite{li2012automatic}. The classification accuracy achieved in detecting polyps with this technique by integrating the features extracted from different sub-image levels compared to a single level is high. By implementing (P) and (R)  with values of 24 and 3, respectively, the classification accuracy in RGB color space achieved is 91.60\%, while in HSI color space it is 91.0\%, showing the performance of technique \cite{li2012automatic}. 
\par 
For detecting polyp inside WCE images, a unique technique termed sparse auto-encoder with image manifold constraint (SSAEIM)  is proposed \cite{yuan2017deep}. Although the shape features implemented in \cite{iwahori2015automatic, bae2015polyp} and (HOG) \cite{dalal2005histograms} are all-inclusive, still they neglect intrinsic information. The discriminative model results in the suggested technique robust and accurate for distinguishing polyps from normal images \cite{yuan2017deep}. SSAEIM for polyp detection was stimulated by sparse autoencoder (SAE), which is an unsupervised learning network for used for learning the features of images automatically.  To avoiding pitfalls such as neglecting intrinsic information, and in order to achieve a discriminative approach, implementation of the image manifold model in combination with theories such as graph and the latest manifold learning is preferred \cite{ma2015local}. For such modification, SAE can be altered by integrating the image manifold results (SAEIM), which ensures similarity of the features in the images of a given category.  Compared to simple neural networks, the deep neural network provides better performance for image characterization. The stacking of several SAEIM layers was employed in the proposed method, in which three SAEIM layers were stacked. Image resizing from $256\times 256$ to $256\times 256$ $\times 3$ was undertaken to reduce the computational complexity, which changed the size of the first layer to 12288 units, followed by 6000 and 1200 units for the second and third layers, respectively. The output layer is the fourth layer comprising 128 units, which produced a discriminative model capable of characterization of the features at a high level. In every layer of SSAEIM, image manifold was applied to maintain high inter-variances and low intra-variances between features .The output of the SSAEIM model involves characterization of each wireless capsule image with dimensions of 128 as a feature, followed by the use of a supervised learner called Softmax classifier for polyp detection. In the model, five parameters (the sparsity penalty control ($\beta$), image manifold constraint($\gamma$), weigh decay cost($\alpha$), learning rate($\mu$), and sparsity parameter ($\rho$)) were set to obtain the best results \cite{yuan2017deep}. Evaluation of the SSAEIM technique was performed with 3000 normal wireless capsule images each having 1000 of the above three categories. The overall recognition accuracy (ORA) of the proposed computational technique for analysis of the wireless capsule images was 98.00\%, while for each category (polyp, clear images, turbid, and bubbles) the accuracy was found to be 98.00\%, 95.50\%, 99.00\%, and 99.50\%, respectively \cite{yuan2017deep}. 
\par 
In contrast with the outdated bag of feature (BoF) technique, an improved bag of features (BoF)  is implemented for classifying polyps inside WCE images \cite{Yuan2016ImprovedBO}. Features such as the complete local binary pattern (CLBP), visual words of the length of 120, $8\times 8$ path size, and SVM are used in line in the proposed technique for automatic detection of polyps in WCE images. The key points, called salient points having useful information is the focus of the proposed technique, around which extraction of the features are done \cite{Yuan2016ImprovedBO}. The SIFT features are extracted due to their rotation and scale in variance \cite{Guo2010ACM}. The five features considered are the SIFT \cite{Lowe2004DistinctiveIF}, LBP \cite{ojala1996comparative}, uniform LBP \cite{ojala2002multiresolution}, CLBP \cite{Guo2010ACM}, and HOG \cite{dalal2005histograms}. Feature integration was performed after acquiring the above textural features by defining the key point as $p_i$ and (x,y) using tactic, where $p_i$ and (x,y) define the pixel and location in the original image, respectively. By selecting a region having a given patch size denoted by $p_i$, the SIFT + LBP descriptor can be generated, represented by the combined vector that defines the whole patch.  
\begin{equation}
SIFT + LBP = [SIFT_i, LBP_i]      
\end{equation}
Conversion of vector patches into a visual of words is done in order to produce a vocabulary by applying K-means to the
data set. After that, the combined features of the testing and training images are coded by assigning the
closest visual words that form a histogram, which later becomes the input for the classifier. The improved BoF technique is tested using two classifiers, i.e., SVM (LibSVM \cite{Chang2011LIBSVMAL} and FLDA \cite{Bishop2007PatternRA}. A total of 2500 WCE images used, of which 2000 were normal and 500 were polyp images. The training is performed with 1000 normal images and 250 polyp images, and the remaining 1000 normal images were used for testing with 250 polyp images \cite{Yuan2016ImprovedBO}. The analysis is performed based on patch sizes of all dimensions for all combinations of features, i.e., [(SIFT+LBP), (SIFT + uniLBP), (SIFT + CLBP and (SIFT +HOG)], and implemented using two classifiers, i.e., SVM and FLDA. The results achieved by using these combinations and classifiers showed that by using SIFT integrated with CLBP and SVM as a classifier, a specificity of 90.88\%, sensitivity of 94.54\%, and accuracy of 93.20\% is  achieved, indicating performance superior to that in recent work done in \cite{li2012automatic}.
\par 
Furthermore, using information based on histogram chromaticity that reflects color and shape information that is further integrated with Zernike moment. This integrated technique attempted to distinguish normal and polyp images inside WCE images. The Zernike moment which is invariant to rotation, scale, and translation is used as a feature extractor in HSI color space.  For the representation of the color features, a two-dimensional histogram in HSI color space is applied for the HS channel; thus, the histogram provided chromaticity information and is called a chromaticity histogram \cite{Li2009IntestinalPR}. Another important factor that impedes recognition (as a low-level feature of the image) is the shape of the polyp. For shape illustration, two techniques can be applied, i.e., contour-based and region-based \cite{Zhang2003EvaluationOM}.  However, due to the background in the capsule images and the fact that is hard to achieve clear and accurate contour, a shape descriptor, which is region-based, was implemented in the proposed technique. In the case where the shape features are invariant to rotation, translation, and scaling, Zernike moments \cite{Teague1980ImageAV} can be applied as they satisfy the mentioned properties. Three-hundred images from the gastrointestinal tract are selected, consisting of 150 images of polyps and 150 normal images. Two classifiers, i.e., the machine learning perception (MLP), having three MLP layers plus two nonlinear outputs, and the support vector machine (SVM) inspired by the work are used. By using Zernike moments with 5 orders, the proposed technique employing the machine learning perception (MLP) classifier resulted, amazingly, in even better results than the support vector machine (SVM), yielding an accuracy, sensitivity, and specificity of 94.20\%, 93.33\%, and 95.07\%, respectively \cite{Li2009IntestinalPR}. 
\par 
Moreover, the detection of colorectal polyps inside WCE images by unsupervised learning is attempted. The proposed technique consists of watershed segmentation employing a unique primary marker selection procedure that is dependent on features such as the Gabor texture and clustering using K-means. To avoid illumination issues, a selection technique termed initial marker dependent on Gabor filters and K-means is implemented \cite{Hwang2010PolypDI}. K-means clustering used to generate clusters having similar properties and the output of the Gabor filters was used to obtain a number of segmented regions based on the elevation levels. Based on these levels, regions chosen as local minima were selected as initial markers. The watershed transform is applied using the selected markers and regions that are more deeply segmented were obtained with the controlled marker. The selected polyp candidates exhibited an elliptical or spherical shape; thus, the curvature center ratio introduced for each region. A total of 128 images used for the proposed technique, consisting of 64 normal and 64 polyp images. An optimal value of “K” chosen,  and the correct polyp segmentation ratio (CPSR) is computed with 100\% sensitivity and 81.00\% specificity \cite{Hwang2010PolypDI}.
An automatic polyp detection based on the decomposition of WCE images via wavelet and curvelet transform is proposed. The proposed detection scheme starts with preprocessing to eliminate environed black regions followed by the data augmentation by applying the geometric technique of rotation, flipping, and cropping \cite{souaidi2018new}. For the multi-resolution analysis, transform like the 2-D discrete wavelet transform, dual-tree complex wavelet transforms (DTCWT), Gabor wavelet transforms, and curvelet transforms is exploited. The distribution of the sub band coefficients is shaped concerning the characteristics of the wavelet-based transform weather it produced a real or complex values coefficient by utilizing Weibull, Generalized Gaussian distribution (GGD), Rayleigh or gamma distributions. Processing of each channel of RGB color space is done to collect all the features to generate a final feature vector. Support Vector Machine classifier (SVM) is used for classifying the images into polyps and normal images.  The data set comprised of 5926 polyp patches and 1864 normal patches after data augmentation. The proposed scheme while using gamma distribution and DTCWT generated a high performance in terms of accuracy, sensitivity, and specificity of 96.00\%, 96.00\%, and 96.00\% respectively \cite{souaidi2018new}.
\par 
As WCE generate a lot of images that result in an imbalance data set. This imbalance of the data set creates a bias towards the non-polyp class in the detection process. A structure for achieving enhanced data sampling is proposed, i.e., Adaboosting, in which unbiased polyp detection is learned. The learning arrangement consists of numerous weak classifiers integrated, having re balanced data sets based on down/up sampling to obtain a polyp detector. Further partial least squares (PLS) is applied in order to improve the capability for feature discrimination among non-polyps and polyps, and high dimensionality \cite{Bae2015PolypDV}. Old techniques based on imbalanced learning are classified into re weighting and re sampling approaches \cite{Galar2012ARO, Hoens2013IMBALANCEDD}. These techniques suffer from a cost sensitivity problem, and to overcome this cost sensitivity boosting (CSB) algorithms such as AdaBoost \cite{Fan1999AdaCostMC} and CSB1/CSB2 \cite{Ting2000ACS} is implemented. Moreover, for slanted data sets, partial least squares (PLS) is used to achieve asymmetric classification \cite{Qu2010AnAC}. Modeling of relationships among groups with latent variables is performed via PLS. The goal of this analysis is to establish latent vectors or orthogonal score vectors by maximization of the covariance among dissimilar variable sets \cite{Hwang2007PolypDI}. The proposed technique uses the latest improvement in the area of imbalanced learning to develop an integrated framework of AdaBoost, re sampling, and partial least squares (PLS), termed as data sampling-based boosting. 
\par 
Despite the enhanced classification achieved with PLS, even with imbalanced datasets, there is still a need to cope with the issue
of imbalance. The synthetic minority over-sampling technique (SMOTE) for the generation of samples in feature space is adopted.  When the data are free of noise, down-sampling is performed with SVM, whereas in the case of noisy data, Tomek links analysis \cite{Kubat1997AddressingTC} is performed. To address the diverse appearance of polyps and imbalanced data sets, a framework design called data sampling-based boosting via integration of suggested feature learning plus data sampling was utilized with algorithms such as AdaBoost.M1 \cite{Polikar2006EnsembleBS}. The entire technique was implemented with 1263 images after checking the data set with CVC-ColonDB \cite{Bernal2012TowardsAP}. Benchmarking of the detection performance using different features (color histogram, LBP, HOG and VHOG) for the same classifier showed that better performance was achieved when VHOG was used in combination with the proposed PLS. Extensive quantitative benchmarking of diverse detectors based on the AdaBoost algorithm using different re sampling approaches employing imbalanced learning techniques was also performed. These detectors include: 
\begin{itemize}
	\item SVM: Single linear support vector machine (SVM) classifier
	\item ESM: Ensemble classifier (No sampling)
	\item ESM-UP: Ensemble classifier (Up-sampling)
	\item ESM-DW: Ensemble classifier (Down-sampling)
	\item ESM-UP-DW: Ensemble classifier (Up/down-sampling)
\end{itemize}
\par 
The results showed better detection performance in terms of a lower miss rate and false positive per image (FFPI), which were
achieved by implementing detectors with the VHOG-PLS feature, using a resampling-based technique, i.e., ESM-UP-DW, irrespective of the size of the dataset and data variations. The cascaded approach \cite{Viola2001RobustRO} was applied for imbalanced datasets via ensemble learning, despite the use of AdaBoost. The proposed technique outperformed the cascaded approach in terms of the recall and precision-recall area under the curve (PR-AUC), yielding higher accuracy and lower miss-detection \cite{Bae2015PolypDV}.
\par 
For detection of polyp inside WCE images, a technique is proposed based on the extraction of local fractal dimension (LFD) features above detected  SIFT key points.  Concatenation of extracted features with Complete local binary pattern (CLBP) or uniform local binary pattern (LBPu) is done in order to assimilate texture information.  Due to the appearance, color, and texture of polyps, extraction of geometrical information was performed by using LFD.  Textural features that are robust to RST are extracted by choosing the CLBP or LBPu from the WCE by concatenation with LFD. The dataset consisted of 2433 capsule images comprising 1700 and 733 normal and polyp images, respectively. The classification is performed with classifiers, such as MLP, SVM, and random forest, in order to differentiate polyp images from normal images \cite{Ansari2017ComputeraidedSF}. For the integration of SIFT + LFD + LBPu and SIFT + LFD + CLBP, the performance achieved with the three classification techniques is assessed, demonstrating the better performance of SVM. The better classification was achieved with SIFT + LFD + LBPu in comparison with SIFT + LFD + CLBP, resulting in precision, accuracy, specificity, and sensitivity of 96.19\%, 97.98\%, 95.86\%, and 100\%, respectively. The area under the curve (AUC) for both combinations (i.e., SIFT + LFD + LBPu for all three classifiers) are computed, and values of 0.9793, 0.9779, and 0.9765 were respectively obtained for SVM, random forest, and MLP, whereas for SIFT + LFD + CLBP with all three classifiers, the AUC was 0.9632, 0.9703, and 0.9632 for SVM, random forest, and MLP, respectively \cite{Ansari2017ComputeraidedSF}.
\par
A modified region-based convolutional neural network (R-CNN) by utilizing Kears deep learning model is proposed for detecting polyps inside WCE images by producing masks around polyps.  The localization of polyps is done by segmentation followed by detection of most likely pixels points of polyps in the mask areas and determining the centroid depicting the location of the polyp \cite{sornapudi2019region}. Data augmentation of WCE images is done by rescaling, rotation, histogram equalization, and gaussian blurring to avoid overfitting during the model.  For the features extraction, ResNet \cite{he2016deep} with Feature Pyramid Network (FPN) \cite{lin2017feature} is implemented that extracts feature from different hierarchical levels with different scales. The Region Proposal Network (RPN) along with feature extraction make the bounding box or proposals where the RPN works in a convolution form by scanning the features received from FPN. The binary masks are generated by assigning the region proposals to various distinct regions of features maps obtained from FPN. These mapped regions are supplied to the RoIAlign module \cite{he2017mask}, supported by convolutional layers and fully connected layers for predicting the location and size of the predicted mask to fit the object. The final binary mask (28 * 28) matches to specific region proposal, and the regions that have a class probability of 0.8 or greater are rated true predictions. For the optimization of the loss function on specific ROI region, stochastic gradient descent (SGD) is used.  Fine-tuning of the model is done on the pre-trained models of COCO weights and Flicker's balloon by using the data set of ImageNet dataset and Microsoft COCO, where the learning rate was kept 0.001, epochs of 1000, and with empirically chosen hyperparameters. The proposed model is implemented for different scales of region proposals such i.e. $8\times 8$, $16\times 16$, $32\times 32$, $64\times 64$, and $128\times 128$ to detect poly inside WCE images of different sizes resulting in precision and F1 score of 98.46\% and 96.10\% respectively \cite{sornapudi2019region}. 

\begin{table}[H]
	\renewcommand{\arraystretch}{0.65}
	\caption{Efficacy of existing methods for Polyp detection in WCE images.}
	\resizebox{11.30cm}{!}{
	\begin{tabular}{|c|c|c|c|c|}
		\hline
		\textbf{Ref No.}  & \textbf{\begin{tabular}[c]{@{}c@{}}Accuracy\\ (\%)\end{tabular}} & \textbf{\begin{tabular}[c]{@{}c@{}}Specificity\\ (\%)\end{tabular}} & \textbf{\begin{tabular}[c]{@{}c@{}}Sensitivity\\ (\%)\end{tabular}} & \textbf{\begin{tabular}[c]{@{}c@{}}Precision\\ (\%)\end{tabular}} \\ \hline
		\textbf{{[}37{]}} & \textbf{NO}                                                      & 67.50                                                               & 100                                                                 & \textbf{NO}                                                       \\ \hline
		\textbf{{[}41{]}} & 98.65                                                            & 98.52                                                               & 98.79                                                               & \textbf{NO}                                                       \\ \hline
		\textbf{{[}43{]}} & 98.43                                                            & 98.23                                                               & 98.67                                                               & \textbf{NO}                                                       \\ \hline
		\textbf{{[}44{]}} & 91.43                                                            & 94.78                                                               & 88.09                                                               & \textbf{NO}                                                       \\ \hline
		\textbf{{[}46{]}} & 91.60                                                            & \textbf{NO}                                                         & \textbf{NO}                                                         & \textbf{NO}                                                       \\ \hline
		\textbf{{[}49{]}} & 98.00                                                            & \textbf{NO}                                                         & \textbf{NO}                                                         & \textbf{NO}                                                       \\ \hline
		\textbf{{[}54{]}} & 93.20                                                            & 90.88                                                               & 95.54                                                               & \textbf{NO}                                                       \\ \hline
		\textbf{{[}59{]}} & 94.20                                                            & 95.07                                                               & 93.33                                                               & \textbf{NO}                                                       \\ \hline
		\textbf{{[}62{]}} & \textbf{NO}                                                      & 81.00                                                               & 100                                                                 & \textbf{NO}                                                       \\ \hline
		\textbf{{[}63{]}} & 96.00                                                            & 96.00                                                               & 96.00                                                               & \textbf{NO}                                                       \\ \hline
		\textbf{{[}64{]}} & \textbf{NO}                                                      & \textbf{NO}                                                         & \textbf{NO}                                                         & 70.67                                                             \\ \hline
		\textbf{{[}75{]}} & 97.98                                                            & 95.86                                                               & 100                                                                 & 96.19                                                             \\ \hline
		\textbf{{[}76{]}} & \textbf{NO}                                                      & \textbf{NO}                                                         & \textbf{NO}                                                         & 98.46                                                             \\ \hline
	\end{tabular}
}
\end{table}
\par 
Table 4 summarizes the results of a survey of the methods for detection of polyps in WCE images. The performance of the methods is represented in terms of the accuracy, specificity, sensitivity, and precision.
\section{Ulcer detection in WCE images}
An automatic technique based on convolutional neural network (CNN) model grounded on Single Short Multi box Detector (SSD) is proposed in to detect erosions and ulceration in WCE images \cite{aoki2019automatic}. For training the model 5360 WCE images having erosions and ulceration collected from 115 patients, while 10,440 free images for testing collected from 64 different patients. In the 10,440 free images, 440 contain erosions while remaining 10,000 images are normal small bowl images. Single Short Multi box Detector (SSD) \cite{liu2016ssd}, is implemented for AI diagnostic system having input in the form of input labeled images via framework of deep learning named Caffe. Stochastic gradient is implemented with 0.0001 learning rate after resizing each image to 300*300.  The annotation of images are done manually indicated by green (true box) and yellow (CNN box) and then based on the condition, if overlapping area among true and CNN box enclosed more than true or CNN box, define it as accurate answer. The second condition is for the existence of multiple CNN boxes in a single image, then detection of an ulceration or erosion is true even 1 boxes accurately detect rather than that all CNN boxes should fulfill criteria. The area under the curve (AUC) for CNN model to detect ulcerations or erosion after confidential interval (CI) adjustment is 95.80\%. The optimal value of cut-off for score of probability according Youden index was 0.481, so areas with score of probability greater than 0.481 are identified as ulcerations or erosions by CNN mode. At this cut-off value, the CNN model accuracy for detection of erosions or ulcerations in terms of accuracy, sensitivity and specificity are 90.81\%, 88.20\% and 90.90\% respectively \cite{aoki2019automatic}.
\par
An automatic ulcer detection system based on color and texture feature extraction is proposed. The proposed system attempt preprocessing by applying median filtering to remove noises from the neighboring pixels \cite{charfi2019computer}. For segmentation purpose, simple linear iterative clustering (SLIC) technique is applied. After the SLIC, color and textural saliency mapping are computed. The color saliency map features are computed based on the mean of the colors of sub-regions of all pixels by utilizing the Euclidean distance. For the textural saliency map features a local binary pattern (LBP) is implemented followed by the fusion process of both maps. Features are extracted by applying Color LBP (CLBP) by estimating the LBP for each channel of RGB color channel. For the spatial layout and local shape of the image, Pyramid of Histograms of Orientation Gradients (PHOG) is employed for extracting descriptors at edges in comparison to HOG. Also, the bag of the visual word is analyzed to extract features by setting the No. of visual words to 500. For classification, every feature extracted from the segmented regions is separately classified by SVM via a radial basis function kernel, MLP and RF classifiers. Contrast to considering the most likely class which observation should fit the label, the generated scores of the classifiers which mean the probability that an observation fit to a particular class. The scores of these classifiers are given as observations to a first-order discrete hidden Markov model (HMM). The proposed detection approach is implemented on two data sets; one having a total of 446 WCE images in which 287 are an ulcer and 159 are normal images. The second data set comprised of a total of 2170 images, out of which 570 are normal and 1600 are ulcer images. The proposed automatic detection system generated a high performance in terms of accuracy, sensitivity, and specificity for both data sets. For the first data set, the accuracy, sensitivity, and specificity achieved is 95.30\%, 96.20\%, and 96.20\% respectively, while for the second data set, the accuracy, sensitivity, and specificity archived is 94.80\%, 96.20\%, and 95.50\% respectively \cite{charfi2019computer}.
\par 
Automatic detection and classification of ulcers proposed for WCE images using deep learning modes of CNN architectures i.e. Alexnet and GoogLeNet. Features such as LBP, color histogram and color coherence are extracted for detailed simulation for checking the performance of CNN model \cite{alaskar2019application}. The CNN layers consisting of neurons arranged in 3 dimensions for converting each 3D input into respective 3D output of activation neuron. The pre-trained models used are fine-tuned through the weight freezing of the initial layer. The pre-trained model GoogLeNet \cite{krizhevsky2012imagenet} in which every layer work as a kernel for extracting finest features is used for object detection and classification. The proposed work use GoogLeNet with addition of four new layers, 50\% dropout probability, fully connected layer (FCN), softmax layer and output layer of classification. The AlexNet \cite{yuan2017deep} used ReLU as an activation function with 50\% dropout probability where feature extractor is used as a first layer plus AlexNet has fewer layers compared to GoogLeNet. Modification is done as addition of one FCN layers, modifying layer 23 to the equal size as the output layer classification and number of classes.  The dataset consist of 1875 images out of which 250 images are normal images while 1525 images are ulcer images and are resized accordingly to train both models. Detailed experimental results are achieved for both models by selecting different learning rates that shows that with learning rate of 0.0001, the classification accuracy reach 100\%, with sensitivity and specificity of 1 for both GoogLeNet and AlexNet. However, AlexNet is outperforming GoogLeNet in the training time by almost 50\% because fewer layers \cite{alaskar2019application}.
\par 
A unique technique for the detection of ulcer and erosion of small intestine is proposed using framework of deep learning for WCE images. 144 patients are selected for the dataset collection, which include 65 normal, 47 ulcer and 32 erosion cases \cite{fan2018computer}. For detection of lesions two independent models i.e. one ulcer and one for erosion detection are trained. For the ulcer detection 3,250 ulcer and 5000 normal images are used, while for erosion detection, 8000 normal and 4,910 erosion images are used. AlexNet \cite{yuan2017deep} is implemented for the detection purpose with both models trained with stochastic gradient descent, 50 batch size, 0.01 of learning rate, 0.9 of momentum and 0.001 of weight decay. The training accuracy after changing six time learning rate for erosion and ulcer model achieved high accuracy level of 94.68\% and 96.36\% respectively. The ROC curve is also evaluated to check the veracity of both models generating AUC value for erosion and ulcer model of 0.9863 and 0.9891 respectively. For further performance demonstration of the technique proposed, comparison with other approached such as SVM and gray scale histogram is done yielding outperforming results. The overall performance for both erosion and ulcer detection model generated high performance values in terms of sensitivity of 93.67\% and 96.80\%, an accuracy of 95.34\% and 95.16\%, a specificity of 95.98\% and 94.79\%, an AUC of 0.9904 and 0.9805 correspondingly \cite{fan2018computer}.  
\par 
For detecting multi-abnormalities inside WCE image, a novel technique consisting of a stage called learning of feature to generate visual words.  From polyps, bleeding, ulcers, and samples of normal WCE images, computation of SIFT feature independently. Segmentation is done to remove the background, highlighting the critical areas of the WCE images for all patches followed by textural and color information extraction via HSC-SIFT \cite{Yuan2017WCEAD}. Then the K-means clustering algorithm is applied on extracted features to get visual words. Integrating these four kinds of visual words used later for classification. At the feature stage coding, a unique saliency and adaptive locality-constrained linear coding (SAALC) is proposed for encoding the images. Encoding of patch features was achieved with SAALC based on adaptive coding computed using the difference in the distance between features and visual words. Compared to traditional coding such as LLC, implementation of adaptive coding bases removes the issue of selecting every patch feature based on static visual words. The max-pooling yield representation of the WCE image was applied after coding of every patch descriptor. For recognition of abnormal images, SVM is used as a classifier \cite{Chang2011LIBSVMAL} using a Gaussian radial basis as a kernel. The data set consists of total 1650 images out of which 500 images each for normal tissues, polyps, and bleeding while rest 150 images are ulcers. Compared to other techniques for multi-abnormality classification, the proposed technique generated the best results, having an overall recognition accuracy (ORA) of 88.61\%. The proposed technique also outperformed the bench marked technique for detecting bleeding, polyps, ulcers, and normal images, with an accuracy of detection for each image type of 96.60\%, 83.50\%, 80.33\%, and 88.20\%, respectively \cite{Yuan2017WCEAD}. 
\par 
In order to detect ulcer in WCE images, a novel textural extraction process is suggested using the multi-resolution property of DCT.  From the sub-bands of DCT, the lacunarity index is computed to obtain textural information \cite{Eid2013ACL}. YCbCr color space was chosen for the analysis, although it was not considered as a feature in this study. DCT is applied on each color channel via wrapping technique in order to remove wavelet weakness, as wavelet cannot adequately signify image singularities like curves and lines while dealing with medical data \cite{Cands2000CurveletsAS}. The parameters (i.e., no. of scales and angles) for DCT analysis are carefully selected, as sub-band information can be lost. Differential lacunarity analysis is used to compute the filling of space while dealing with a large homogeneous data set. In spite of using the traditional gliding box algorithm (GBA) for computing the lacunarity, that is only suited for binary images, the technique used to implement the differential lacunarity analysis (LAC) suitable for gray scale images \cite{Cands2000CurveletsAS}. By choosing a fixed box size of “r=3" and a window size of “w= 4-$box_m$", where “$box_m$" is equivalent to the minimum aspect of every sub-image the LAC was calculated. As this computation can result in a high dimensional vector feature, leading to degradation of the accuracy and performance speed, the hyperbolic function is implemented to avoid such issues, represented as \cite{allain1991characterizing}:
\begin{equation}
{  \Lambda = \frac{b}{w^a} + c   }       
\end{equation}
\par
For classification, SVM having the kernel of radial basis function with the scalar unit factor is implemented. Ten-fold cross-validation is applied for testing and training of the classifier. The experimental process is performed with 130 normal and 130 abnormal images. The results of using different angles and color channels are evaluated, demonstrating that the “Cr” channel gave good results, whereby choosing  an optimal angle in the “Cr” channel by trying different combinations led to an accuracy of 86.54\%, the specificity of 88.56\%, and sensitivity of 84.51\% \cite{Eid2013ACL}.
\par 
In order to classify normal and abnormal images inside WCE, a color wavelet transform (CWC) based on DWT is proposed.  Textural color and analysis are done due to variance in the appearance of ulcer and color of blood. For multi resolution and textural analysis, three-level DWT is used. Four second-order statistical models (correlation, entropy, angular the second moment, and inverse difference) are used at different orientations ($0^{\circ}$,$45^{\circ}$, $90^{\circ}$, and $135^{\circ}$) to provide discrimination with a high level of accuracy \cite{Liu2012ANA}. Seventy-two CWC vectors were computed after calculating the resemblance among the wavelet features in order to avoid waste of computational time due to repeated information. Human textural perception defined by Texton \cite{Julesz1981TextonsTE} is specifically used for categorization and is symbolized by the texton map “T". For a pair of textural layout filters defined by (r, T) where “r" is the region of the image and “t" is the texton, at location “i", the response of the feature is proportional to the pixels for the region of offset “r+1" with the texton index “t", represented as:
\par
A modified joint boost algorithm \cite{Torralba2004SharingVF} is implemented for iterative selection of the highest discriminative textural layout as a“weak learner’ for every texton. This selection is then integrated into the classifier, allowing every single weak learner to be used for classification of a number of classes at one time \cite{Torralba2004SharingVF}. The data set collected from Jin shan corporation consisted of 100 images, out of which 45 images are abnormal tissues. For training, 45 samples randomly opted while 45 are for testing. The remaining 10 samples are used for validation. The experimental analysis is divided into two phases comprised of selecting the best color space in order to define the best textural properties, followed by bench marking relative to techniques such as co-occurrence matrix (CM) and wavelet transform (WT).The results achieved with the proposed technique indicated better performance in terms of the sensitivity 82.30\% and specificity 89.10\% \cite{Liu2012ANA}. 
\par 
Furthermore, textural feature analysis is done for the detection of ulcers inside WCE images using curvelet-based pattern i.e. LBP \cite{li2009texture}. LBP shows resistance to illumination variances and it can discriminate small structures such as dark and bright spots. The features extracted by LBP are the mean, standard deviation, energy, kurtosis, skew, and entropy. The multi-resolution analysis is performed using wavelet theory as it yields smooth functions when dealing with one-dimensional data.  To capture more features in more directions, curvelet transforms is applied, which intimates a curve in the shape followed by a precise scaling rule \cite{do2003finite}. The data set collected from 20 patients consisted of 3600 images, out of which 1800 normal and 1800 ulcer images. To prevent over-fitting, 4-fold cross-validation was used in this specific technique. SVM and MLP neural network are used to verify the performance of the extracted features for distinguishing normal from ulcerated regions. For performance analysis, the proposed technique with suggested CLBP features is bench marked in “RGB" and “YC{b}C{r}" color space using different types of features, i.e., LBP features, CWC, and CM. Implementation of MLP neural network in “YC{b}C{r}" color space for the specific data set yield better performance, with an accuracy, sensitivity, and specificity of 92.37\%, 93.28\%, and 91.46\%, respectively \cite{li2009texture}.
\par 
Moreover, detection of ulcer inside the WCE image is done by using local feature integration. The proposed method is based on the bag-of-words (BOW) model and feature fusion method. Implementation of LBP and SIFT as feature extractor done on each patch after patching scheme. Clustering for every patch the feature which resulted in a vocabulary dictionary is performed with the K-means clustering algorithm \cite{yu2012ulcer}. In the proposed technique, the spatial pyramid kernel is used to obtain spatial pyramid BOW histograms to consider spatial layout information \cite{lazebnik2006beyond}. The main addition in this feature is an increment of the coarser grid over the space of the image at 1....L resolution so that the grid at level 1 has 2l-1 cells beside every dimension. A total of 344 WCE images comprising 172 normal and 172 ulcer images were utilized. Additionally, 120 ulcer images and 120 normal images were chosen for testing. For classification, SVM is implemented for training these features.  For feature fusion, linear classifier dependency modeling (LCDM) based on Bayesian theory is used to avoid large intra-class and small inter-class variations \cite{ma2011linear}. For extraction of local regions for the BOW model, techniques such as grid and Harris affine are also used, followed by the construction of vocabularies of different sizes for pyramid kernel. The results showed that by using the best classifier with feature fusion in combination with LBP produced a higher accuracy, sensitivity, and specificity of 89.58\%, 99.17\%, and 80.00\%, respectively than those of the feature fusion in combination with SIFT \cite{yu2012ulcer}. 
\par 
To detect ulcers in WCE images, a two-stage fully automatic technique is proposed. In the first stage, a saliency detection technique based on multi-level super pixel illustration was proposed for shaping the ulcer. Salient regions that are semantically significant and perceptible were found via image segmentation using multilevel super-pixel. The analogous saliency was estimated according to the textural and color features for every level, as these features are related to different primary areas of super-pixel sizes. A final saliency is generated by fusing all levels of saliency maps. The second stage involved the characterization of the images from the saliency map in order to efficiently encode the features for ulcer detection via locality constrained linear coding (LLC) \cite{yuan2015saliency}. The suggested saliency detection technique comprised three major steps, starting with multi-level segmentation using simple linear iterative clustering (SLIC), which segment the WCE images from rough to a fine level into numerous super-pixels. Thereafter, saliency regions are detected based on textural and color contrast for every level of super-pixel. The Leung-Malik (LM) filter bank is used for extraction of the textural features, as LM is a multi resolution, multi-orientation, and multi-scale filter comprising 48 filters \cite{leung2001representing}.
\begin{equation}
\label{tariq}
texture_s = \frac{1}{L} \sum_{l-1} ^{L} texture_{s^l}
\end{equation}
\par 
For color-based saliency detection, numerous trial, and error experiments are performed in different color spaces to preserve
useful information and HSV and CMYK are found to be optimal for ulcer recognition. The mean and variance values from HSV (i.e.,S plane) and for CMYK (i.e., M) are selected for generating the color feature matrix. Regions with variations in the color distribution are chosen to obtain color saliency. Thereafter, by taking the mean of these different levels, the final saliency is calculated as:
\begin{equation}
color_s = \frac{1}{L} \sum_{l-1} ^{L} color_{s^l}
\end{equation}
\par 
Finally, a multi-level texture and color super-pixel saliency is proposed by fusion of the saliency regions. The final saliency map was obtained as shown below:
\begin{equation}
final_s = texture_s \odot color_s \odot K
\end{equation}
\par 
Where “$\odot$" is the Hadamard product matrix representing  Gaussian kernel focused at the center of the image which slowly
declines towards the edges in order to simulate human perception \cite{yuan2015saliency}. In the second stage, the modified locality constrained linear coding (LLC) technique is used to combine the acquired saliency map with the max-pooling technique for WCE images. It consists of three steps, starting with the extraction of local descriptors (dense scale-invariant feature transform dSIFT \cite{lowe2004distinctive}, a dense histogram of oriented gradients (dHOG) \cite{dalal2005histograms}, and dense uniform local binary patterns (duniLBP) \cite{ojala2002multiresolution} for each image independently to generate a code book or visual vocabulary via the K-means clustering technique. After obtaining three codebooks, the mapping of each descriptor matrix is performed with code books for image illustration. Initially, the original LLC was used to encode every single descriptor obtained for each testing and training image. After coding, all codes were pooled together to obtain a representation of the WCE image. The images are finally represented by the concatenation of the results generated from three descriptors as:
\begin{equation}
final_Y = [dSIFT_{y}T  \quad dHOG_{y} T   \quad duniLBP{y}T ]^T
\end{equation}
\par 
The dataset for the experimental analysis consisted of 170 normal and 170 ulcer images taken from 20 patients. 1225 patches are extracted, having dimensions of $16\times 16$ with three kinds of features, i.e., dSIFT), dHOG and duniLBP. SVM having a Gaussian radial basis function is then applied as a classifier. For testing and training, 20\% and 80\% of the images are respectively used, with 5-fold cross-validation. Tuning of the vocabulary size (M) for the proposed technique is carefully executed, and the results generated with M=50 are the best, with mean specificity, sensitivity, and accuracy values of 91.18\%, 94.12\%, and 92.65\%, respectively indicating the validity of the proposed algorithm \cite{yuan2015saliency}.
\par 
Table 5 summarizes the results of a survey of the methods for detection of ulcer in WCE images. The performance of the methods is represented in terms of the accuracy, specificity, sensitivity, and precision.

\begin{table}[H]
	\renewcommand{\arraystretch}{0.65}
	\caption{Efficacy of existing methods for ulcer detection in WCE images.}
	\resizebox{11.30cm}{!}{
	\begin{tabular}{|c|c|c|c|c|}
		\hline
		\textbf{Ref No.}  & \textbf{\begin{tabular}[c]{@{}c@{}}Accuracy\\ (\%)\end{tabular}} & \textbf{\begin{tabular}[c]{@{}c@{}}Specificity\\ (\%)\end{tabular}} & \textbf{\begin{tabular}[c]{@{}c@{}}Sensitivity\\ (\%)\end{tabular}} & \textbf{\begin{tabular}[c]{@{}c@{}}Precision\\ (\%)\end{tabular}} \\ \hline
		\textbf{{[}80{]}} & 90.81                                                            & 90.90                                                               & 88.20                                                               & \textbf{NO}                                                       \\ \hline
		\textbf{{[}82{]}} & 94.80                                                            & 95.50                                                               & 96.20                                                               & \textbf{NO}                                                       \\ \hline
		\textbf{{[}83{]}} & 100                                                              & 100                                                                 & 100                                                                 & \textbf{NO}                                                       \\ \hline
		\textbf{{[}85{]}} & 95.16                                                            & 94.79                                                               & 96.80                                                               & \textbf{NO}                                                       \\ \hline
		\textbf{{[}86{]}} & 88.61                                                            & \textbf{NO}                                                         & \textbf{NO}                                                         & \textbf{NO}                                                       \\ \hline
		\textbf{{[}87{]}} & 86.54                                                            & 88.56                                                               & 84.51                                                               & \textbf{NO}                                                       \\ \hline
		\textbf{{[}90{]}} & \textbf{NO}                                                      & 89.10                                                               & 83.20                                                               & \textbf{NO}                                                       \\ \hline
		\textbf{{[}93{]}} & 92.37                                                            & 91.46                                                               & 93.28                                                               & \textbf{NO}                                                       \\ \hline
		\textbf{{[}95{]}} & 89.58                                                            & 80.00                                                               & 99.17                                                               & \textbf{NO}                                                       \\ \hline
		\textbf{{[}98{]}} & 92.65                                                            & 91.18                                                               & 94.12                                                               & \textbf{NO}                                                       \\ \hline
	\end{tabular}
}
\end{table}

\section{Bleeding detection in WCE images}
Bleeding inside the gastrointestinal (GI) track is an important factor that need to be detected timely. However, the images generated from WCE are numerous and having different luminance quality at different places, making a hectic task for physicians to diagnose.  This bleeding inside GI track can lead to further harmful diseases, such as tumor, poly and ulcer etc. making the necessity of techniques to detect efficiently bleeding inside WCE.
\par
To detect bleeding inside a frame of WCE videos, a color, and textural features extraction technique is proposed. Color analysis is done for removing dark and bright blocks to compute luminance by taking the square root of a sum of single RGB blocks \cite{pogorelov2019bleeding}. Edges detection and removal for avoiding false results using the canny operator followed by morphological dilation which dilates the information detected. Color features are extracted by computing the red ratio for single pixels in the R channel of the RGB color space. The extracted features based on color is fed to SVM that uses three types of kernel i.e. radial basis function (RBF), polynomial, and linear for classification purpose. By applying a statistical approach of computing textural features, the gray‐level co‐occurrence matrix (GLCM) is utilized that extracted 22 textural features.  A classification approach based on color and texture integration is implemented for the discrimination of bleeding and non-bleeding pixels. Random Tree (RT), Random Forest (RF), and Logistic Model Tree (LMT) classifiers are used having different combinations of color and texture features as an input vector. The data set was collected by endoscopists from Endoscopy unit at University of Malay Medical Center (UMMC), Kuala‐Lumpur, Malaysia consisting of 1200 frames out of which 800 are bleeding while 400 are non-bleeding frames. For the frame level of bleeding detection, SVM using RBF as a kernel resulted in high accuracy, specificity, sensitivity, and F1 score, of 97.77\%, 95.55\%, 97.76\%, and 97.80\% respectively. For the entire pixel-level bleeding detection the proposed approach of integrated features classification generated high results in terms of accuracy, specificity, sensitivity, and F1 score, of 97.76\%, 95.59\%, 97.76\%, and 97.76\% respectively \cite{pogorelov2019bleeding}.
\par 
A unique approach based on textural feature extraction is proposed in which classification is achieved by using Probabilistic Neural Network (PNN). For detection of bleeding areas, an intelligent programming technique that are tested on image level and pixel level \cite{pan2011bleeding}. Inside WCE images, non-bleeding areas are dissimilar bleeding regions even in colored or gray images. Color textural feature extraction is done for RGB and HSI color spaces making 6D color feature vector x= (R, G, B, H, S, I). The PNN used as a classifier is feed forward three layered neural network that include, input layer, radial basis layer, and competitive layer. The color feature vector is the input for PNN classifier, which recognize m=1 and m=0 as bleeding and non-bleeding pattern respectively. After that, a bleeding detection technique is applied after removing dark pixels from WCE image followed by forming a feature vector of each pixel as: x= (R, G, B, H, S, I). A pixel by pixel scanning is done on WCE image and recognition is done by PNN for bleeding and non-bleeding pattern. The software for bleeding detection is test both for pixel and image level. For pixels level 930,575 pixels containing 339,850 bleeding and 590,725 non-bleeding pixels are manually detected. For image level, 3,172 bleeding images and 11,458 non-bleeding images are manually detected.  The results generated by image level testing gives a higher performance values in terms of sensitivity and specificity of 93.10\% and 85.60\% respectively compared to pixel level testing as it’s not necessary for whole image level to be so sensitive for pixel level of minor bleeding \cite{pan2011bleeding}.
\par 
Furthermore, a novel technique is proposed for detection of bleeding inside WCE video that is based on pixel grouping via pixel segmentation \cite{fu2013computer}. Except edge pixels, non-bleeding and bleeding pixels mostly have similar luminance and vary in redness resulting to apply canny detector in L channel. The canny edge detector detects the pixels of edge while pixels of bleeding can be survived followed by morphological operation to dilate the image. Due to high luminance of WCE images, pixel to pixel segmentation is a time consuming and computational cost task, therefore adaptive grouping of pixels based on location and color via super pixel segmentation is performed. The proposed super pixel segmentation is the modified approach from \cite{levinshtein2009turbopixels,achanta2012slic} introducing a cluster center as an initial seed growth that depend on similarity of pixel for finding a cluster rather than evolution of curve, making it faster than \cite{fu2013computer}. Furthermore, Gaussian filter is applied to remove noise from WCE images and to find connectivity for bleeding region, new label assigning is done on disjoint segment resulting a single super pixel for even a small bleeding region.  A similarity measure among color similarity and spatial distance is computed based on Euclidean distance and spatial position of pixel respectively as shown by,
\begin{equation}
{  S = S_{color} + \lambda\times S_{spatial}}
\end{equation}
{The $\lambda$ controls the two factors in above equation. }
\par 
Implementing SVM \cite{chang2011libsvm} for training, detecting patterns of bleeding and non-bleeding is done for super pixel to super pixel of WCE image. Color is the main feature while bleeding detection and SVM is trained with RGB color. Training data set for SVM constitute of 60,000 feature vectors comprising of 20,000 bleeding pixels and 40,000 pixels of non-bleeding. The whole data set comprised of 5,000 WCE image, containing 4,000 non-bleeding and 1,000 bleeding images collected from 20 patients. Different radial basis of SVM are implemented such as RBF, Linear and Poly. The results achieved using RBF as radial basis for SVM generate a higher accuracy, specificity and sensitivity having values of 95.00\%, 94.00\% and 99.00\% respectively \cite{fu2013computer}.
\par 
Moreover, examination and detection of bleeding in RGB normalized color space is done for WCE images. The proposed technique is grounded on variation profile of inter plane intensity that extract effective region of interest (ROI) \cite{kundu2018automatic}. Normally, the region surrounding in WCE images are having unnecessary information and analysis in RGB color space have certain limitation due to the high correlation of channels making the detection of bleeding hard. For remedy, this study is proposing a normalized color scheme represented by (rgb) having range of (0, 1) after transformation of coordinates of each pixel \cite{cheng2001color, andreadis1990image}. A pre-processing step for ROI extraction is done to have more discriminative features that indicate only bleeding areas rather entire image pixels consideration. The ROI identification is most crucial step in the proposed study that is done on normalized plane. Based on linear separable boundary, a condition is made in order to distinguish between bleeding and non-bleeding areas as given by,
\begin{equation}
{ r(i,j)\geq n\times g(i,j) }
\end{equation}
\par 
Here, “n" indicate a slope of discriminative line r-g plane. The pixels fulfilling the condition are considered to be ROI. Segmentation of ROI is done to provide certain of bleeding image for testing yet not assuring that exact bleeding region, leading to extract features from ROI for further classification. Features extraction is done at the ROI via histogram from values of pixel of normalized planes of WCE image as it can reproduce the areas of bleeding even if it is small or large in some bins. From the intensities of pixels of the corresponding plane of g, histograms are generated followed bin frequency values suggestions as bleeding region features. The features extracted are inputted to the K-nearest neighbor (KNN) classifier to classify non-bleeding and bleeding images. A distance function is calculated among test and train data sets that become test data set for the classifier. After bleeding image detection, marking the bleeding region automatically is helpful for physicians to identify the area. Morphological operation of two steps is done in order to detect bleeding in the ROI frames selected and classified. The proposed method outperformed other techniques and generated detection accuracy, sensitivity and specificity of 97.86\%, 95.20\% and 98.32\% respectively \cite{kundu2018automatic}.
\par
Furthermore, a color histogram of block statistics, namely CHOBS approach is proposed to detect bleeding inside WCE video. Preprocessing is done in the first step of the analysis to remove black pixels of edges and corner areas. The CHOBS extract block-based local feature from each color channel offering better feature representation in contrast to single pixel-based feature \cite{ghosh2018chobs}. By integrating local block of three channels of RGB color space, an index value is determined.  For obtaining the index value of per pixel, the binary values of L most significant bits (MSBs) of the individual color plane are used. From the extracted local features, extraction of color histogram-based global features is done. A color histogram, which is extracted from those index values, presents a distinct color texture feature. To select bin with the higher color histogram and information for bleeding images, feature selection and dimensionality reduction are done. For the former, histogram pattern is utilized while for the latter principal component analysis (PCA) is applied. To classify the non-bleeding and bleeding WCE images, K-nearest neighbor (KNN) classier is employed. The proposed technique also localize the bleeding region in addition to the detection of bleeding by classifying all blocks of an image into non-bleeding and bleeding classes. Then recognizing the label of every pixel within a block for bleeding or non-bleeding followed by morphological operation fine-tuning. The dataset consisted of 2350 WCE images out of which 1900 non-bleeding and 450 bleeding images. For the detection of the bleeding frame, the proposed technique generated an accuracy, specificity, and sensitivity of 97.85\%, 99.15\%, and 99.47\% respectively\cite{ghosh2018chobs}.
\par 
Another work that is extending Bag of Words technique \cite{sivic2003video, csurka2004visual} taking advantage of color feature extraction to classify bleeding images from non-bleeding images is proposed \cite{yuan2015bleeding}. Avoiding whole image analysis, region of interest (ROI) is outlined by considering extreme square inscribed in the circular image representing main image features. Color features extraction is done in “YCbCr", “RGB", “HSV" and “LAB" via opting 10\% normal images and 10\% bleeding images. Using k-means clustering algorithm \cite{kanungo2002efficient} that get input represented image pixel vectors in the color space, consistent cluster centers are calculated. The cluster centers that are concatenated generated from normal WCE data set and bleeding data set become vocabulary of visual words. Mapping 3D color data on each image point to closest visual words and computing the amount of each visual word resulting a histogram (w, d). In K-size clusters, wi indicate the ith visual word while di count occurring frequency. SVM \cite{chang2011libsvm} and KNN used as a classifier for discriminating bleeding form normal images. For localization of bleeding area, a two-stage saliency extraction strategy is applied. The first stage saliency comprises of different weight assignments particularly in CMYK and CIELAB color space. The second stage is in “RGB" color space, where bleeding region is defined by R value threshold. Regions similar to value of red colors are considered to be bleeding region which are then integrated for all three channels saliency maps. At the end, the two saliencies are fused together having bleeding localization information with different aspects, and is given as:
\begin{equation}
{ S_{final} = w_{1} \times S_{stage1}+w_{2}\times S_{stage2} }
\end{equation}
Where $S_{final}$ is the fused saliency map $w_{1}+w_{2}=1$ The data set consisted of 2400 WCE images obtained from 10 different patients, out of which 2000 are normal images and 400 are bleeding images. The proposed technique generated best results for setting cluster centers 80 having accuracy, sensitivity and specificity of 95.75\%, 92.00\% and 96.50\% respectively \cite{yuan2015bleeding}.
\par 
Moreover, an efficient extraction of features scheme is suggested to detect bleeding frame in WCE video automatically. Pre-processing is done in order to remove the black pixels in the outer regions of WCE images having low intensities resulting desired shape in the form of circular image. As red and green channels greatly influence the bleeding and non-bleeding regions, so instead of red (R), green (G), and blue (B) analysis independently a transform plane is constructed based on ratio of pixel intensity if red and green (R/G) planes \cite{ghosh2015block}. Due to illumination variation in WCE images, analysis of single pixel can lead to distortion and computational burden, and extracted features might not be offer reliable characteristics. So extraction of features is done based on block pixels as the R/G ration is high for red (R) for bleeding pixels. The study suggested an R/B domain maximum value of block pixels for extracting potential features. For specific color spatial distribution, block feature histogram is computed after construction of R/G plane followed by feature extraction for each block.  For bleeding images the patterns of histogram vary significantly from non-bleeding images. For classification, the K-nearest neighbor (KNN) is use a function based on Euclidean distance to calculate pattern in the features in the test image and K neighboring patterns in the data set of training. The data set consisted of 1000 WCE images, out of which 800 are non-bleeding and 200 are bleeding images. The proposed technique after setting K=10 for testing achieved accuracy, sensitivity and specificity of 96.10\%, 96.48\% and 96.01\% respectively \cite{ghosh2015block}.
\par 
A semantic segmentation with combination of CNN approach is implemented for detection of gastrointestinal angiectasia (GIA) inside small-bowl (SB) capsule endoscopy \cite{leenhardt2019neural}. The SB capsule endoscopic frames are collected from Computer Assisted Diagnosis for Capsule Endoscopy database (CAD-CAP) containing 6,360 still frames. Manual annotation of 6,360 still frames are done, in which extraction of 2,946 of still frames having vascular lesion. The non-hemorrhagic still frames are chosen by experts same as for vascular lesion. Two data sets are made i.e. A1 and A2 that are used for training and testing purpose respectively. Another two data sets i.e. N1 and N2 comprising of 20,000 normal still frames are extracted which is used for training and testing purpose collected from SB-CEs database. Initially morphological features such as color, shape, contour and size are extracted by designing a handcrafted colorimetric segmentation method based on image saturation variation and red over green ratio. For more satisfactory results, CNN is implemented as it extracts first the finest features and then classification is done based extraction of features.  An acceptable distinction during training and learning phase is necessary among GIA still frames by using dataset NI as a control for A1. Similarly, for N2 and A2 for testing phase. Two data sets are created after identification of 600 non-hemorrhagic GIA splitting into 300 still frames each (A1 and A2) from 2,946 still frames having vascular lesions. Similarly, 600 normal still frames are chosen for creating control data sets i.e. (N1 and N2). The proposed technique of segmentation with a CNN model for extraction of deep features generated a specificity and sensitivity of 96.00\% and 100\% for detection of GIA respectively \cite{leenhardt2019neural}.
\par 
Furthermore, an automatic scheme is implemented to detect Angiodysplasis lesions by categorizing latent region of interest and then classifying them by combination features such as color-based, statistical, texture and morphological operation. To solve the issue of unbalanced sampling among non-pathological and pathological regions, a boosted decision tree classification technique is opted. Image pre-processing is done by applying a contrast-limited adaptive Histogram equalization (CLAHE) for enhancing the local contrast among intestinal background and an angiodysplasia regions followed by application of “RGB" de-correlation stretching for detecting local heterogeneity \cite{noya2017automated}. Potential region selection is started by analyzing that green channel have high color contrast among background and lesion, so normalized threshold of 0.15 is selected to generate a mask which act as kernel for segmentation scheme. For homogeneous regions, hole filling algorithm which is morphological reconstruction is applied followed by erosion process and finally division of regions into smaller regions which are reflected as initial candidate regions. Filtering for the initial nominee regions is done heuristically by determining a threshold of heuristic for the region, potential degree of regions and perimeter via their geometric properties. To find weather the region is a lesion or not, the degree of region is the ratio among pixels in the area and circumscribed rectangle pixels. Splitting of ROI is done to avoid the addition of non-lesion pixels by using intensities variance in each channel defining as a scale for region heterogeneity. The features extracted for an angiodysplasia lesion are color, texture and geometric and the proposed work extracted 11 statistical, 8 geometrical and 5 texture features. The features ae extracted in 4 different color spaces which are “YCbCr, “RGB", “HSV" and “CIElab" resulting 200 total features. A step wise regression analysis to avoid feature redundancy which reduce features set to 80.  Classification of latent ROI is done as 97.70\% region is recognized as non-pathological regions by the ROI detection scheme leading to unbalanced data set. To alleviate this effect, RUSBoosted \cite{seiffert2009rusboost} trees classifier is implemented yielding results in the form of receiver operating characteristic (ROC) and area under the curves (AUC). Total data set consisted of 45,664 and 1,029 non-pathological and pathological samples respectively for training and testing in which 22,832 and 514 regions are non-lesion and with lesion respectively. Different classifiers with 10-fold cross validation are evaluated for unbalance data sets yielding high accuracy, still the RUSBoosted approach in comparison generated much more higher accuracy of 96.80\% with AUC of 98.69\% \cite{noya2017automated}.
\par 
An automatic detection for obscure bleeding using a novel color feature extraction method is presented using support vector machine (SVM) as a classifier \cite{liu2009obscure}. The three matrices represented by M1, M2, and M3 are used to store three channels intensities of RGB color space. Rather than extracting features from whole WCE image, down sampling is done by block division phenomena that have only ROI followed by pixel transformation into vectors called “Raw Vectors”. The proposed feature selection is based on the ratio of intensities between red and green as M1/M2 for bleeding as blue channel (M3) have no such influence for bleeding region. SVM from \cite{chang2001libsvm} with different function as kernel is used for classification purpose having input fed from extracted feature step. The data set consisted of 800 images out of which 400 are normal and 400 are bleeding images. The proposed technique using ratio features generated sensitivity and specificity more than 99\% compared to raw vectors and color histogram features. Overall the using linear kernel for SVM, a high average specificity is achieved while for RBF as a kernel resulted in high average sensitivity compared to other kernels \cite{liu2009obscure}.
\par
Furthermore, a universal extraction of feature based technique is proposed that use a transform color domain analysis to discriminate bleeding region in WCE images. Pre-processing is done to remove the dark sides of WCE images leaving only the central portion. Due to illumination variation and high correlation among RGB channels, bleeding discrimination become hard from normal images as it share the same color property \cite{ghosh2018automatic}. So a narrow band histogram is constructed on the pixel intensity ration of G/R as blue color is less significant for bleeding case. Features are extracted on G/R ratio plane overcoming the illumination problems. Statistical features of order higher and lower are extracted from that transformed color domain. The potential features extracted for this study included median, variance, and kurtosis.  A count based feature from pixel intensity ratio is done, that make a threshold value $T_{G}/R$ for discriminating bleeding and non-bleeding region. Based on the threshold, the pixel intensity for bleeding should have G/R intensity of pixel ratio less than $T_{G}/R$. For classification, supervised with linear kernel and Sequential minimization optimization (SMO) is used for binary classification. Additionally, a detection approach is provided for WCE videos using a post processing process from the results obtained from classifier. An investigation zone is formed based on the analysis that bleeding and non-bleeding frames occur continuously in consecutive (L) frames. Frames are identified and using edge detection algorithm like Sobel operators, edges are removed followed by same transformation and ratio of G/R which is here 0.5 for bleeding. Morphological dilation and erosion is applied to find the connectivity and remove small regions.  The data set comprised of 2350 WCE images obtained from 20 different videos, out of which 1900 are non-bleeding while 450 are bleeding images. The proposed technique using SVM with linear fucntion as kernel generated a high sensitivity, accuracy and specificity of 97.75\%, 97.96\% and 97.99\% respectively \cite{ghosh2018automatic}.
\par 
Based on segmentation and choosing channels of color for feature analysis having a lot of information for bleeding regions, a novel technique is proposed for bleeding detection. The color space conversion and selection of channel is considered as a pre-processing. The hue, RGB and CIE lab are chosen for specific feature analysis where by a naïve approach for every value of pixel in each channel, an index is considered making a look up table (LUT) \cite{Hajabdollahi2018SegmentationOB}. Increased in the LUT content in their corresponding indices can be seen for bleeding regions in the data set. By this approach, saturation, gray scale and “a” are chosen superlative color features. A multi layer perception (MLP) artificial neural network (ANN) with hidden layer as 40-20-8 is implemented on three selected color channels for classification. Patches having dimensions of $5\times5$ are extracted from three colors spaces becoming input for MLP where central pixel class during training phase is reflected as network output. The sigmoid function is used for binary decision and later quantized by weights adjustment. The network is trained and tested with 50 images with 5 fold cross validation and results are achieved in terms of DICE of high average value of 84.03\% \cite{Hajabdollahi2018SegmentationOB}.

\par 
A novel approach for automatic bleeding detection in WCE images assimilating features like handcrafted (HC) and convolutional neural network (CNN). The proposed technique efficiently discriminate bleeding region from non-bleeding even on small training data set. The technique comprised of three steps, extraction of features, combination of these features and finally classification process \cite{jia2017gastrointestinal}. For HC features, k-means clustering is applied to get information in the form of 2-D histogram for each image. Obtaining HC feature vector in 1-D for each WCE input image by saving values of y-axis that  represent pixel number fitting to each  cluster. Lab color space is suited for such feature extraction having length of 50 for HC feature vector \cite{jia2016gi}. Feature extraction via CNN \cite{krizhevsky2012imagenet} is done by making 8 layered CNN model comprising of 3 conv layers , 3 pooling layers and 2 fully connected layers with leaning rate, momentum and weight decay of 0.01, 0.9 and 0.004 respectively. The data set comprised of 1500 WCE images obtained from 80 different patients, out of which 1200 are normal images while 300 are bleeding images having dimension of 240*240*3. The proposed integrated approach resulted in higher values of precision, recall and F1 score of 94.79\%, 91.00\% and 92.85\% compared to other feature extraction based technique \cite{jia2017gastrointestinal}.
\par 
Table 6 summarizes the results of a survey of the methods for detection of bleeding in WCE images. The performance of the methods is represented in terms of the accuracy, specificity, sensitivity, and precision.
\begin{table}[H]
	\renewcommand{\arraystretch}{0.65}
	\caption{Efficacy of existing methods for Bleeding detection in WCE images.}
	\resizebox{11.30cm}{!}{
		\begin{tabular}{|c|c|c|c|c|}
			\hline
			\textbf{Ref No.}   & \textbf{\begin{tabular}[c]{@{}c@{}}Accuracy\\ (\%)\end{tabular}} & \textbf{\begin{tabular}[c]{@{}c@{}}Specificity\\ (\%)\end{tabular}} & \textbf{\begin{tabular}[c]{@{}c@{}}Sensitivity\\ (\%)\end{tabular}} & \textbf{\begin{tabular}[c]{@{}c@{}}Precision\\ (\%)\end{tabular}} \\ \hline
			\textbf{{[}101{]}} & 97.76                                                            & 95.59                                                               & 97.76                                                               & 97.76                                                             \\ \hline
			\textbf{{[}102{]}} & \textbf{NO}                                                      & 85.60                                                               & 93.10                                                               & \textbf{NO}                                                       \\ \hline
			\textbf{{[}103{]}} & 95.00                                                            & 94.00                                                               & 99.00                                                               & \textbf{NO}                                                       \\ \hline
			\textbf{{[}107{]}} & 97.86                                                            & 98.32                                                               & 95.20                                                               & \textbf{NO}                                                       \\ \hline
			\textbf{{[}110{]}} & 97.85                                                            & 99.15                                                               & 99.47                                                               & 95.75                                                             \\ \hline
			\textbf{{[}113{]}} & 95.75                                                            & 896.50                                                              & 93.10                                                               & 95.24                                                             \\ \hline
			\textbf{{[}115{]}} & 96.10                                                            & 96.01                                                               & 96.48                                                               & \textbf{NO}                                                       \\ \hline
			\textbf{{[}116{]}} & \textbf{NO}                                                      & 96.00                                                               & 100                                                                 & \textbf{NO}                                                       \\ \hline
			\textbf{{[}117{]}} & 96.80                                                            & 96.80                                                               & 89.51                                                               & \textbf{NO}                                                       \\ \hline
			\textbf{{[}119{]}} & $>99$                                                            & $>99$                                                               & $>99$                                                               & \textbf{NO}                                                       \\ \hline
			\textbf{{[}121{]}} & 97.96                                                            & 97.99                                                               & 97.75                                                               & 88.20                                                             \\ \hline
			\textbf{{[}122{]}} & \textbf{NO}                                                      & \textbf{NO}                                                         & \textbf{NO}                                                         & 98.00                                                             \\ \hline
			\textbf{{[}123{]}} & \textbf{NO}                                                      & \textbf{NO}                                                         & \textbf{NO}                                                         & 94.79                                                             \\ \hline
		\end{tabular}
	}
\end{table}
\par 
Furthermore, detailed information is provided in Table 7, describing the specifications and source of the collected data set. Thus, we have provided a comprehensive details with in-depth probing of the methods underlying data set collection and the number of frames selected. Details specifying division of the data set into testing and training to conduct experimental analysis is also been provided, and lastly, cross-validation information to avoid over-fitting is presented.
\par 
Figure 3 shows a comprehensive overview of how each anomaly is detected in WCE images based on the technique implemented. The techniques for tumor, polyp, and ulcer detection are divided into machine learning (ML), and ML is further divided into SVM and deep learning; if the method is not a machine-learned approach, it is categorized as “other techniques".
\newpage
\begin{table}[H]
	\renewcommand{\arraystretch}{1.20}
	\caption{Specifications of the collected data set.}
	\resizebox{12.00cm}{!}{
	\begin{tabular}{|c|c|c|c|c|c|}
		\hline
		\textbf{Ref No.}   & \textbf{\begin{tabular}[c]{@{}c@{}}Data Set \\ origin\end{tabular}}                                                                                                                                                                       & \textbf{\begin{tabular}[c]{@{}c@{}}No. of frames\\ in Data set\end{tabular}}           & \textbf{\begin{tabular}[c]{@{}c@{}}Testing \\ data\end{tabular}}                      & \textbf{\begin{tabular}[c]{@{}c@{}}Training \\ data\end{tabular}}                     & \textbf{\begin{tabular}[c]{@{}c@{}}Cross \\ Validations\end{tabular}} \\ \hline
		\textbf{{[}18{]}}  & \textbf{ShariatiHospital in Tehran, Iran(M2A Micro capsule)}                                                                                                                                                                              & \textbf{315}                                                                           & \textbf{120}                                                                          & \textbf{110}                                                                          & \textbf{Unspecified}                                                  \\ \hline
		\textbf{{[}19{]}}  & \textbf{Capucho’s Hospital in Lisbon}                                                                                                                                                                                                     & \textbf{3,936}                                                                         & \textbf{Unspecified}                                                                  & \textbf{Unspecified}                                                                  & \textbf{10 fold}                                                      \\ \hline
		\textbf{{[}20{]}}  & \textbf{Obtained by four gastroenterologists(Capsule data)}                                                                                                                                                                               & \textbf{1,800}                                                                         & \textbf{Unspecified}                                                                  & \textbf{Unspecified}                                                                  & \textbf{10 fold}                                                      \\ \hline
		\textbf{{[}21{]}}  & \textbf{Obtained by experts(Given Imaging CE)}                                                                                                                                                                                            & \textbf{1,200}                                                                         & \textbf{900}                                                                          & \textbf{300}                                                                          & \textbf{4 fold}                                                       \\ \hline
		\textbf{{[}22{]}}  & \textbf{Olympus CF-100 HLendoscope}                                                                                                                                                                                                       & \textbf{1,380}                                                                         & \textbf{1,200}                                                                        & \textbf{180}                                                                          & \textbf{Unspecified}                                                  \\ \hline
		\textbf{{[}23{]}}  & \textbf{PillCam Companyhttp://www.givenimaging.com/}                                                                                                                                                                                      & \textbf{800}                                                                           & \textbf{200}                                                                          & \textbf{600}                                                                          & \textbf{Unspecified}                                                  \\ \hline
		\textbf{{[}24{]}}  & \textbf{PillCam SB2}                                                                                                                                                                                                                      & \textbf{300}                                                                           & \textbf{Unspecified}                                                                  & \textbf{Unspecified}                                                                  & \textbf{3 fold}                                                       \\ \hline
		\textbf{{[}25{]}}  & \textbf{PillCam SB2 J. Lau and Y. Chan, two experts in Prince of Wales Hospital, Hong Kong}                                                                                                                                               & \textbf{1,200}                                                                         & \textbf{60}                                                                           & \textbf{540}                                                                          & \textbf{10 fold}                                                      \\ \hline
		\textbf{{[}28{]}}  & \textbf{Hospital dos Capuchosin Lisbon by Doctor Jaime Ramos (CE data)}                                                                                                                                                                   & \textbf{400}                                                                           & \textbf{200}                                                                          & \textbf{200}                                                                          & \textbf{Unspecified}                                                  \\ \hline
		\textbf{{[}29{]}}  & \textbf{Unspecified}                                                                                                                                                                                                                      & \textbf{12, 090}                                                                       & \textbf{20\%(2418)}                                                                   & \textbf{80\%(9672)}                                                                   & \textbf{Unspecified}                                                  \\ \hline
		\textbf{{[}30{]}}  & \textbf{Unspecified}                                                                                                                                                                                                                      & \textbf{Unspecified}                                                                   & \textbf{Unspecified}                                                                  & \textbf{Unspecified}                                                                  & \textbf{Unspecified}                                                  \\ \hline
		\textbf{{[}31{]}}  & \textbf{\begin{tabular}[c]{@{}c@{}}Dr. Hossein Asl Soleimani and Shariati Hospital, Iran\\ (M2A capsule endoscopy device, Given Imaging Company,)\end{tabular}}                                                                           & \textbf{60}                                                                            & \textbf{12}                                                                           & \textbf{48}                                                                           & \textbf{Unspecified}                                                  \\ \hline
		\textbf{{[}33{]}}  & \textbf{Hospital dos Capuchos in Lisbon by Doctor Jaime Ramos}                                                                                                                                                                            & \textbf{396}                                                                           & \textbf{192}                                                                          & \textbf{204}                                                                          & \textbf{Unspecified}                                                  \\ \hline
		\textbf{{[}34{]}}  & \textbf{$Pillcam^R$ SB, device by Given Imaging}                                                                                                                                                                                          & \textbf{435}                                                                           & \textbf{15}                                                                           & \textbf{70}                                                                           & \textbf{15 fold}                                                      \\ \hline
		\textbf{{[}35{]}}  & \textbf{$Pillcam^R$ SB, device byGiven Imaging}                                                                                                                                                                                           & \textbf{1,385}                                                                         & \textbf{15}                                                                           & \textbf{70}                                                                           & \textbf{15 fold}                                                      \\ \hline
		\textbf{{[}36{]}}  & \textbf{Capucho’s Hospital in Lisbon}                                                                                                                                                                                                     & \textbf{3200}                                                                          & \textbf{Unspecified}                                                                  & \textbf{Unspecified}                                                                  & \textbf{10 fold}                                                      \\ \hline
		\textbf{{[}37{]}}  & \textbf{Digestive Specialists, Dayton, OH}                                                                                                                                                                                                & \textbf{50}                                                                            & \textbf{Unspecified}                                                                  & \textbf{Unspecified}                                                                  & \textbf{Unspecified}                                                  \\ \hline
		\textbf{{[}41{]}}  & \textbf{\begin{tabular}[c]{@{}c@{}}(http://www.depeca.uah.es/colonoscopy\_dataset/) \\ and  (https://polyp.grand-challenge.org/databases/)\end{tabular}}                                                                                  & \textbf{14,000}                                                                        & \textbf{7,000}                                                                        & \textbf{7,000}                                                                        & \textbf{Unspecified}                                                  \\ \hline
		\textbf{{[}43{]}}  & \textbf{\begin{tabular}[c]{@{}c@{}}Department of Electronics, University of Alcala \\ (http://www.depeca.uah.es/colonoscopy\_dataset/),\\  and Endoscopic Vision Challenge\\ (https://polyp.grand-challenge.org/databases/)\end{tabular}} & \textbf{$>14,000$}                                                                     & \textbf{4200}                                                                         & \textbf{9800}                                                                         & \textbf{10 fold}                                                      \\ \hline
		\textbf{{[}44{]}}  & \textbf{Unspecified}                                                                                                                                                                                                                      & \textbf{872}                                                                           & \textbf{272}                                                                          & \textbf{600}                                                                          & \textbf{Unspecified}                                                  \\ \hline
		\textbf{{[}46{]}}  & \textbf{M2A WCE, Given Imaging Company}                                                                                                                                                                                                   & \textbf{2,500}                                                                         & \textbf{1250}                                                                         & \textbf{1250}                                                                         & \textbf{10 fold}                                                      \\ \hline
		\textbf{{[}49{]}}  & \textbf{James Lau in Prince of Wales Hospital in Hong Kong,Pillcam SB WCE system}                                                                                                                                                         & \textbf{3,000}                                                                         & \textbf{Unspecified}                                                                  & \textbf{Unspecified}                                                                  & \textbf{3 fold}                                                       \\ \hline
		\textbf{{[}54{]}}  & \textbf{Qilu Hospital, Shandong UniversityM2A WCE, Given Imaging Company}                                                                                                                                                                 & \textbf{2,500}                                                                         & \textbf{1250}                                                                         & \textbf{1250}                                                                         & \textbf{Unspecified}                                                  \\ \hline
		\textbf{{[}59{]}}  & \textbf{James Lau in Prince of Wales Hospital in Hong Kong, Pillcam SB WCE system}                                                                                                                                                        & \textbf{300}                                                                           & \textbf{Unspecified}                                                                  & \textbf{Unspecified}                                                                  & \textbf{3 fold}                                                       \\ \hline
		\textbf{{[}62{]}}  & \textbf{Unspecified}                                                                                                                                                                                                                      & 128                                                                                    & \textbf{Unspecified}                                                                  & \textbf{Unspecified}                                                                  & \textbf{Unspecified}                                                  \\ \hline
		\textbf{{[}63{]}}  & \textbf{World Endoscopy Organization web site}                                                                                                                                                                                            & \textbf{5,926}                                                                         & \textbf{1,186}                                                                        & \textbf{4,740}                                                                        & \textbf{10 fold}                                                      \\ \hline
		\textbf{{[}64{]}}  & \textbf{CVC-ColonDB and Augmented Dataset}                                                                                                                                                                                                & \textbf{1,263}                                                                         & \textbf{Augmented Data set}                                                           & \textbf{CVC-ColonDB}                                                                  & \textbf{5 fold}                                                       \\ \hline
		\textbf{{[}75{]}}  & \textbf{www.capsuleendoscopy.org}                                                                                                                                                                                                         & \textbf{2,433}                                                                         & \textbf{1283}                                                                         & \textbf{1150}                                                                         & \textbf{Unspecified}                                                  \\ \hline
		\textbf{{[}76{]}}  & \textbf{Mayo Clinic(PillCam SB3)}                                                                                                                                                                                                         & \textbf{1,800}                                                                         & \textbf{55}                                                                           & \textbf{429}                                                                          & \textbf{Unspecified}                                                  \\ \hline
		\textbf{{[}80{]}}  & \textbf{\begin{tabular}[c]{@{}c@{}}Ethics committee of The University of Tokyo (no. 11931) \\ and Japan Medical Association (ID JMA-IIA00283)\\ Pillcam SB2 or SB3 WCE device (Given Imaging, Yoqneam, Israel).\end{tabular}}             & \textbf{15,800}                                                                        & \textbf{10,440}                                                                       & \textbf{5,360}                                                                        & \textbf{Unspecified}                                                  \\ \hline
		\textbf{{[}82{]}}  & \textbf{\begin{tabular}[c]{@{}c@{}}Data set 1: W.E. Organization: ‘WEO clinical endoscopy atlas’\\ Data set 2: C. Endoscopy: ‘Capsule endoscopy products’\end{tabular}}                                                                   & \textbf{\begin{tabular}[c]{@{}c@{}}Data set 1 (446)\\ Data set 2 (21700)\end{tabular}} & \textbf{\begin{tabular}[c]{@{}c@{}}Data set 1 (235)\\ Data set 2 (1120)\end{tabular}} & \textbf{\begin{tabular}[c]{@{}c@{}}Data set 1 (211)\\ Data set 2 (1050)\end{tabular}} & \textbf{3 fold}                                                       \\ \hline
		\textbf{{[}83{]}}  & \textbf{Dr. Khoroo’sMedical Clinic/Trust. Available online:http://www.drkhuroo.in/\#}                                                                                                                                                     & \textbf{1,875}                                                                         & \textbf{105}                                                                          & \textbf{421}                                                                          & \textbf{Unspecified}                                                  \\ \hline
		\textbf{{[}85{]}}  & \textbf{Unspecified}                                                                                                                                                                                                                      & \textbf{8,250}                                                                         & \textbf{500}                                                                          & \textbf{2,000}                                                                        & \textbf{Unspecified}                                                  \\ \hline
		\textbf{{[}86{]}}  & \textbf{\begin{tabular}[c]{@{}c@{}}Pillcam SB WCE system and the \\ Rapid Reader 6.0 software (both from Given Imaging Co., Ltd.)\end{tabular}}                                                                                           & \textbf{1,650}                                                                         & \textbf{\begin{tabular}[c]{@{}c@{}}20\%\\ (330)\end{tabular}}                         & \textbf{\begin{tabular}[c]{@{}c@{}}80\%\\ (1,320)\end{tabular}}                       & \textbf{5 fold}                                                       \\ \hline
		\textbf{{[}87{]}}  & \textbf{Unspecified}                                                                                                                                                                                                                      & \textbf{260}                                                                           & \textbf{Unspecified}                                                                  & \textbf{Unspecified}                                                                  & \textbf{10 fold}                                                      \\ \hline
		\textbf{{[}90{]}}  & \textbf{\begin{tabular}[c]{@{}c@{}}Jin Shan Corporation (Yang Qi, Tang Du Hospital \\ of  Fourth Military Medical University)\end{tabular}}                                                                                               & \textbf{100}                                                                           & \textbf{45}                                                                           & \textbf{45}                                                                           & \textbf{10 fold}                                                      \\ \hline
		\textbf{{[}93{]}}  & \textbf{James Lau in Prince of Wales Hospital in Hong Kong, Pillcam SB WCE system}                                                                                                                                                        & \textbf{3,600}                                                                         & \textbf{Unspecified}                                                                  & \textbf{Unspecified}                                                                  & \textbf{4 fold}                                                       \\ \hline
		\textbf{{[}95{]}}  & \textbf{Given  Imaging Ltd.}                                                                                                                                                                                                              & \textbf{584}                                                                           & \textbf{240}                                                                          & \textbf{344}                                                                          & \textbf{Unspecified}                                                  \\ \hline
		\textbf{{[}98{]}}  & \textbf{\begin{tabular}[c]{@{}c@{}}Pillcam SB WCE system and the \\ Rapid Reader 6.0 software (both from Given Imaging Co., Ltd.)\end{tabular}}                                                                                           & \textbf{340}                                                                           & \textbf{\begin{tabular}[c]{@{}c@{}}20\%\\ (68)\end{tabular}}                          & \textbf{\begin{tabular}[c]{@{}c@{}}80\%\\ (272)\end{tabular}}                         & \textbf{5 fold}                                                       \\ \hline
		\textbf{{[}101{]}} & \textbf{University of Malay Medical Center (UMMC), Kuala‐Lumpur, Malaysia}                                                                                                                                                                & \textbf{1200}                                                                          & \textbf{700}                                                                          & \textbf{500}                                                                          & \textbf{10 fold}                                                      \\ \hline
		\textbf{{[}102{]}} & \textbf{Given  Imaging Ltd.}                                                                                                                                                                                                              & \textbf{14,630}                                                                        & \textbf{Unspecified}                                                                  & \textbf{Unspecified}                                                                  & \textbf{Unspecified}                                                  \\ \hline
		\textbf{{[}103{]}} & \textbf{Unspecified}                                                                                                                                                                                                                      & \textbf{5,000}                                                                         & \textbf{Unspecified}                                                                  & \textbf{Unspecified}                                                                  & \textbf{10 fold}                                                      \\ \hline
		\textbf{{[}107{]}} & \textbf{The capsule endoscopy website,” 2014, http://www.capsuleendoscopy.org}                                                                                                                                                            & \textbf{2,300}                                                                         & \textbf{Unspecified}                                                                  & \textbf{Unspecified}                                                                  & \textbf{10 fold}                                                      \\ \hline
		\textbf{{[}110{]}} & \textbf{The Capsule Endoscopy.  Available: http://www.capsuleendoscopy.org}                                                                                                                                                               & \textbf{2350}                                                                          & \textbf{Unspecified}                                                                  & \textbf{Unspecified}                                                                  & \textbf{10 fold}                                                      \\ \hline
		\textbf{{[}113{]}} & \textbf{Unspecified}                                                                                                                                                                                                                      & \textbf{2,400}                                                                         & \textbf{Unspecified}                                                                  & \textbf{Unspecified}                                                                  & \textbf{10 fold}                                                      \\ \hline
		\textbf{{[}115{]}} & \textbf{The capsule endoscopy website,” 2014, http://www.capsuleendoscopy.org}                                                                                                                                                            & \textbf{1,000}                                                                         & \textbf{Unspecified}                                                                  & \textbf{Unspecified}                                                                  & \textbf{Unspecified}                                                  \\ \hline
		\textbf{{[}116{]}} & \textbf{CAD-CAP is a French national multicenter database. SB-CE videos.}                                                                                                                                                                 & \textbf{2,946}                                                                         & \textbf{300}                                                                          & \textbf{300}                                                                          & \textbf{Unspecified}                                                  \\ \hline
		\textbf{{[}117{]}} & \textbf{Given Imaging wireless capsules Pillcam SB 2and Pillcam SB 3.}                                                                                                                                                                    & \textbf{1,648}                                                                         & \textbf{824}                                                                          & \textbf{824}                                                                          & \textbf{10 fold}                                                      \\ \hline
		\textbf{{[}119{]}} & \textbf{Unspecified}                                                                                                                                                                                                                      & \textbf{800}                                                                           & \textbf{20\%(160)}                                                                    & \textbf{80\%(640)}                                                                    & \textbf{Unspecified}                                                  \\ \hline
		\textbf{{[}121{]}} & \textbf{TThe capsule endoscopy website,” 2014, http://www.capsuleendoscopy.org}                                                                                                                                                           & \textbf{2,350}                                                                         & \textbf{20\%(235)}                                                                    & \textbf{90\%(2,115)}                                                                  & \textbf{10 fold}                                                      \\ \hline
		\textbf{{[}122{]}} & \textbf{https: //sites.google .com/site/farahdeeba073/Research/resources}                                                                                                                                                                 & \textbf{100}                                                                           & \textbf{50}                                                                           & \textbf{50}                                                                           & \textbf{5 fold}                                                       \\ \hline
		\textbf{{[}123{]}} & \textbf{Unspecified}                                                                                                                                                                                                                      & \textbf{1,500}                                                                         & \textbf{400}                                                                          & \textbf{1000}                                                                         & \textbf{10 fold}                                                      \\ \hline
	\end{tabular}
}
\end{table}
\par 

\begin{figure}[H]
	\centering
	\includegraphics[width=5.20in,height=18.50in,keepaspectratio]{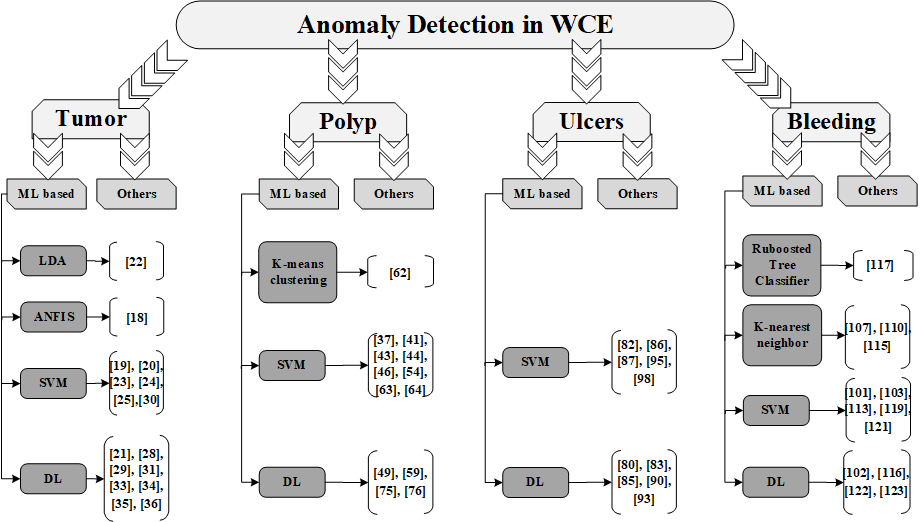}
	\caption{{Detailed information of anomaly detection in WCE images.}}
	\label{Fig2}
\end{figure}
\section{Possible future work: Proposed Solution}
Based on the above studies done, we came up with possible future work by enlisting some of the key problems related to the images generated by wireless capsule endoscopy. The motivation behind for providing a possible future work is to offer a solution to such key issues.
\subsection{Motivation}
Primarily wireless capsule endoscopy is hectic and time consuming task and the output generated should be refined up to the perceptual quality of physicians to diagnose the diseases timely. Firstly, the output generated from WCE is compressed frame due to the limitation of battery life and storage capacity of WCE device resulting in the degradation of the quality of images. Secondly, transmitting WCE frames to a remotely placed physicians can also degrade the quality of the images as a communication channel are vulnerable to noises. Finally, the technique implemented previously either focusing the conventional way of detecting (tumor, polyp, and ulcers) comprising of pre-processing step followed by specific classifiers. Moreover,  mostly the WCE frames while transmission to remote physicians gets affected by additive white Gaussian noise (AWGN), and compression artifacts, etc. leading the necessity of denoising approach before the frames are classified into normal and abnormal frames. To cope with such issues, we proposed a cascade approach of a neural network consisting of two neural networks as explained below. 
\subsection{Proposed approach for joint classification of tumors, polyps, and ulcers}
The proposed approach for the classification of tumors, polyps, and ulcers comprise a cascaded approach of neural network where two neural networks in a feed-forward way are implemented. To the best of knowledge, the proposed cascaded approach is haven’t been implemented nor suggested for the classification of GI diseases when dealing with WCE data. 
\par
The proposed cascaded approach is illustrated in figure 4 where it can be seen that there are two neural networks, i.e. denoising convolutional neural network (DnCNN) \cite{zhang2017beyond} and convolutional neural network (CNN) . 

\begin{figure}[t]
	\centering
	\includegraphics[width=4.80in,height=6.0in,keepaspectratio]{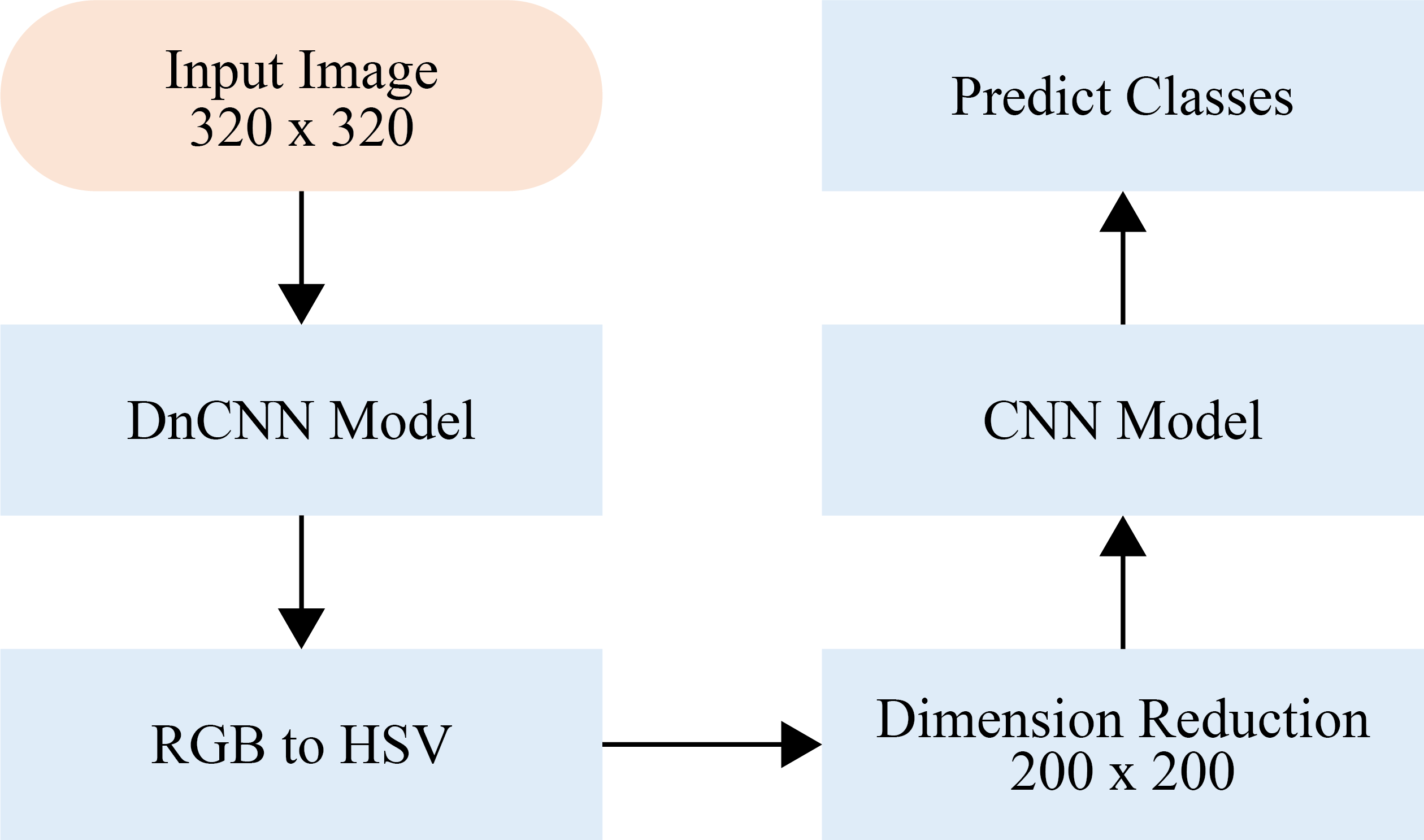}
	\caption{{Proposed Cascaded Model for joint classification of tumors, polyps, and ulcers. }}
	\label{Fig2}
\end{figure}

It can be seen that the proposed cascaded model for joint classification of tumors, polyps, and ulcers has input that is generated from WCE having dimensions of $320\times320$ as shown in figure 4. The WCE frames during the generation phase get compressed and while transmitting over a communication channel to a remote physician gets degraded in terms of perceived quality leading to a time-consuming task for diagnosis. Therefore, the WCE images is passed through a machine-learned denoiser DnCNN that consists of training the network with noisy WCE (tumor, polyps, and ulcers) images. The concept of implementing DnCNN as a residual learner to compute the amount of noise and subtracting that noise from noisy model by a feed-forward convolutional neural network (CNN). To achieve this, a feed-forward convolutional network is fed with a WCE (tumor, polyps, and ulcers) images. The architecture of DnCNN comprises of three layers i.e. Conv+ReLU, Conv+BN+ReLU, and Conv with a depth of “D" as shown in figure 5. For the first layer of the network Conv+ReLU, 64 filters having a size of $3\times 3 \times c$  are implemented to produce 64 feature maps; for non-linearity, rectified linear units (ReLU max (0, .)) are used. Here {“c"} shows the number of channels, three in this case, concerning the colored WCE (tumor, polyps, and ulcers) of the network. For layer 2, Conv+BN+ReLU with 64 filters size of $3\times 3 \times 64$  is chosen with the inclusion of batch normalization between convolution and ReLU. For reconstructing the WCE (tumor, polyps, and ulcers) images, the final layer with Conv having $3\times 3 \times 64$ is used inside the network \cite{zhang2017beyond}.
\begin{figure}[H]
	\centering
	\includegraphics[width=5.10in,height=10.0in,keepaspectratio]{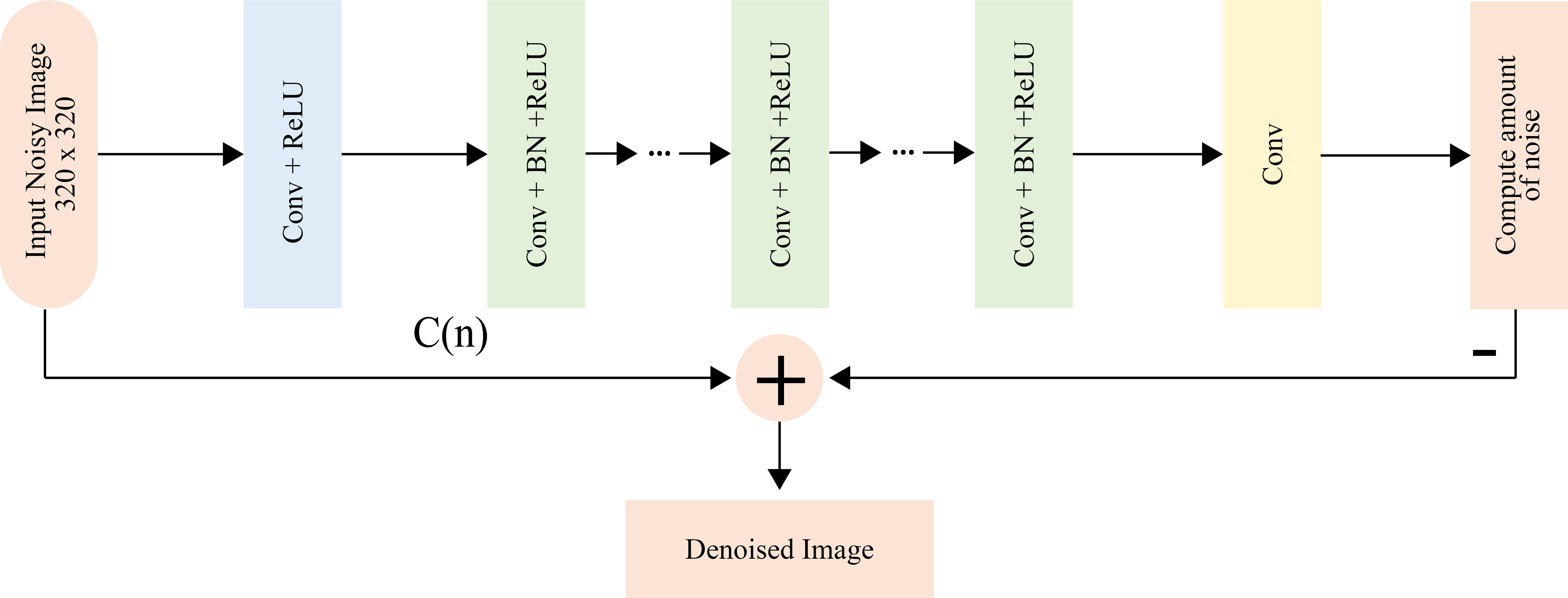}
	\caption{Denoising convolutional neural network for noisy WCE images}
	\label{Fig2}
\end{figure}
\par 
Here the DnCNN model gets an input in the form of noisy WCE (tumor, polyps, and ulcers) images as $y=x+v$, where adoption of residual learning is done to train $R(y)=v$ as a residual mapping generating $x=y-R(y)$. The averaged mean squared error between the desired residual mages and estimated ones from the noisy input can be adopted as losses in the DnCNN’s function to learn the trainable parameters “$\theta$" \cite{zhang2017beyond}.

\begin{equation}
{ l(\theta) = \frac{1}{2N} \sum_{i=1}^{N} || R(y_{i} ; \theta) - (y_{i} - x_{i}) ||_{R}^{2} }
\end{equation}
{ where $ \{   (y_i, x_i ) \}_{i=1}^{N} $ indicates “N" noisy-clean training patch (images) pairs. }
\par 
The denoised images obtained from DnCNN operation are then used for the classification purpose; starting with the color conversion from RGB to HSV, as HSV colorspace refects better visual quality perception for medical imaging analysis. The images dimension are reduced from $320\times320$ to $200\times200$ that are to be fed to CNN model \cite{waqas2018good}. The network architecture is illustrated in figure 6, comprising of 4 layers (L1 - L4) of convolution, 2 layers of max pooling having drop out of 0.2, 1 dense layer having activation function (Softmax) and 1 output layer having three neuron for generating output as (tumor, polyp, and ulcer). For the first layer (L1), input image with dimension $200\times200$ will be passed having filter size of 32 with ReLU as an activation function. The same configuration will be used for L2 reducing the dimension of image to $198\times198$ followed by Max pooling layer of size $2\times2$ with drop out of 0.2. For L3 and L4 layers filter size increased to 64 with ReLU as an activation function, and then again Max pooling will be applied to get features. Flattening will be done in order to have single continuous linear vector that is connected to dense layer. There will be three neurons in the final output layer that will classify diseases based on  category of 0, 1, 2 for tumor, polyp, and ulcer respectively. Furthermore explanation of both networks are the contents of our next research work. 
\begin{figure}[t]
	\centering
	\includegraphics[width=4.80in,height=8.0in,keepaspectratio]{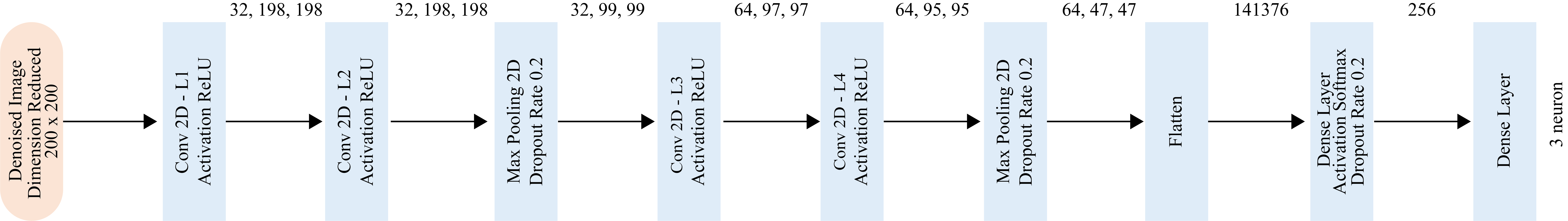}
	\caption{{Convolution neural network (CNN) for classification of (tumor, polyp, and ulcer). }}
	\label{Fig2}
\end{figure}

\section{Formal Discussion}
This section comprises of a formal discussion throwing light on the approaches made to detect various anomalies found in WCE images. The formal discussion section also encompasses the benchmarking of our possible proposed solution with the approaches been adopted for each anomaly detection. 
\par 
For tumor section, the techniques adopted to detect tumor inside WCE images are mostly based on SVM \cite{liu2016detection, li2011computer,faghih2016singular, li2009small, li2012tumor, Ashokkumar2014AutomaticDO} and deep learning (DL) \cite{alizadeh2017detection, karkanis2003computer, barbosa2009automatic, li2017convolutional, Mohapatra2012LymphocyteIS,  Barbosa2008DetectionOS, Sindhu2017AutomaticDO,  Sindhu2017ANM} after extracting features. Preprocessing step for each technique is an essential step followed by a classification approach. For instance, in \cite{li2012tumor} LBP is used as a texture feature extractor with two more feature extractor i.e.  SVM-SFFS and SVM-RFE on 1200 images yielding good performance. Similarly, in \cite{li2017convolutional} the data set is large, implementation of CNN models like LeNet, etc., and high-performance results. Albeit, each technique is following a generalized rule of preprocessing followed by a classification approach for classifying only one disease at a time, none of the studies are attempting to address the issues such as JPEG compression during the generation of WCE videos and AWGN effects while transmitting these frames to a remote physician.  The proposed possible future work is concentrating on the cascaded approach for denoising these artifacts via DnCNN \cite{zhang2017beyond} and then proposing deep learning model \cite{waqas2018good} that will categorize tumor, polyp, and ulcer in a joint classification manner. 
\par 
Similarly for polyp section, in the proposed survey paper are based on SVM \cite{Karargyris2011DetectionOS, Billah2017AnAG, Yuan2014ANF, li2012automatic, Yuan2016ImprovedBO, Ansari2017ComputeraidedSF} and DL  \cite{yuan2017deep, Li2009IntestinalPR, Bae2015PolypDV} where features are extracted and those features are used as an input for classification for each approach.  Take the case in \cite{Karargyris2011DetectionOS} where SUSAN and Gabor filters are used for edge detection process in HSV color spaces. Then SVM as supervised learning is implemented for the classification of polyp and ulcer as the data set consisted of both diseases. Also in \cite{ Yuan2014ANF}, M-LBP is used as a feature extractor for spatial locality followed by SVM. Furthermore, for the DL approach, the Zernike moment analysis is done for feature extraction in \cite{Li2009IntestinalPR} in HSI color space followed by MLP as a  classifier. Although each method is following a generalized rule of preprocessing accompanied by a classification method for classifying only one disease at a time, none of the studies are striving to address the concerns such as JPEG compression during the generation of WCE videos and AWGN effects while transmitting these frames to a remote physician.  The suggested possible future work is focusing on the cascaded approach for denoising these artifacts via DnCNN \cite{zhang2017beyond} and then proposing deep learning model \cite{waqas2018good} that will categorize tumor, polyp, and ulcer in a joint classification manner.
\par 
For ulcer detection research in specific survey comprises of SVM \cite{Yuan2017WCEAD, Eid2013ACL, yu2012ulcer, yuan2015saliency}  and DL  \cite{aoki2019automatic, alaskar2019application, fan2018computer, li2009texture} approaches. The attempt made as in \cite{Yuan2017WCEAD} to use HSC-SIFT for segmentation and textural and color information extraction followed by SVM. Moreover, a DL approach in \cite{li2009texture} for ulcer detection using LBP for textural analysis followed by the MLP neural network for classification purpose. The same justification is provided as given above for the comparison of the proposed possible solution with work done previously.  
\par 
Furthermore, efforts are made to reflect the performance analysis of each disease and section for their respective technique in the survey along with performance metrics and with their references, as shown in table 2, 3, 4, 5, and 6. Table 7 illustrates in detail the source of data set collection along with training and testing of images with cross-validation to show the trade-off between the number of images with the approach being made along with its performance.    

\section{Conclusion}
This paper presents a comprehensive survey of the state-of-the-art computer-aided methods that are used to identify, detect and classify anomalies in the GI tract of living beings capture through WCE process. Several anomalies occur in the GI tract by this review paper only encompasses published methods related to Polyp, Ulcer, and Tumor. A classification based on learning-based methods and other methods is provided as well for each of the three anomalies.
\par 
This paper aims to provide the readers a single platform to gain access to the recently published articles related to computer aided detection of diseases in WCE videos. According to the authors’ extensive review and the works presented in this paper, there has been no work that provides joint classification of all the three mentioned anomalies. At the end of the paper, we have provided a proposal for joint classification of polyp, tumor, and ulcer in WCE videos, where a cascaded approach of implementing neural networks is proposed.

\section*{Acknowledgment}
This work was supported by Priority Research Centers Program through the National Research Foundation of Korea(NRF) funded by the Ministry of Education, Science and Technology”(2018R1A6A1A03024003).

\bibliography{mybibfile}

\end{document}